\documentclass[pdflatex]{sn-jnl}
\usepackage[utf8]{inputenc}
\usepackage{amsfonts}
\usepackage{amsmath}

\usepackage{array}
\usepackage{makecell}
\newlength\savedwidth
\newcolumntype{?}{!{\vrule width2pt}}
\newcommand\thickhline{\noalign{\global\savedwidth\arrayrulewidth\global\arrayrulewidth 2pt}
\hline
\noalign{\global\arrayrulewidth\savedwidth}}

\usepackage{subcaption}

\DeclareMathAlphabet\mathbfcal{OMS}{cmsy}{b}{n}

\usepackage[table]{xcolor}

\usepackage{array}
\usepackage{makecell}
\usepackage{multirow}

\newcolumntype{+}{!{\vrule width 2pt}}

\usepackage{lineno}

\title{Network geometry of the Drosophila brain}

\author[1]{Bendegúz Sulyok}
\author[2]{Sámuel G. Balogh}
\author[3,1,*]{Gergely Palla}

\affil[1]{Dept. of Biological Physics, Eötvös Loránd University, H-1117 Budapest, Pázmány P. stny. 1/A, Hungary}
\affil[2]{
Faculty of Electrical Engineering, Mathematics and Computer Science, Delft University of Technology, 2600 GA Delft, The Netherlands}
\affil[3]{Semmelweis University, Faculty of Health and Public Administration, Health Services Management Training Centre,  H-1125, Kútvölgyi út 2, Budapest, Hungary}

\affil[*]{gergely.palla@emk.semmelweis.hu}

\abstract{
The recent reconstruction of the Drosophila brain provides a neural network of unprecedented size and level of details. In this work, we study the geometrical properties of this system by applying network embedding techniques to the graph of synaptic connections. Since previous analysis have revealed an inhomogeneous degree distribution, we first employ a hyperbolic embedding approach that maps the neural network onto a point cloud in the two-dimensional hyperbolic space. In general, hyperbolic embedding methods exploit the exponentially growing volume of hyperbolic space with increasing distance from the origin, allowing for an approximately uniform spatial distribution of nodes even in scale-free, small-world networks. By evaluating multiple embedding quality metrics, we find that the network structure is well captured by the resulting two-dimensional hyperbolic embedding, and in fact is more congruent with this representation than with the original neuron coordinates in three-dimensional Euclidean space. In order to examine the network geometry in a broader context, we also apply the well-known Euclidean network embedding approach Node2vec, where the dimension of the embedding space, $d$ can be set arbitrarily. In 3 dimensions, the Euclidean embedding of the network yields lower quality scores compared to the original neuron coordinates. However, as a function of the embedding dimension the scores show an improving tendency, surpassing the level of the 2d hyperbolic embedding roughly at $d=16$, and reaching a maximum around $d=64$. Since network embeddings can serve as valuable inputs for a variety of downstream machine learning tasks, our results offer new perspectives on the structure and representation of this recently revealed and biologically significant neural network.
}

\begin{document}

\flushbottom
\maketitle
%
%
\thispagestyle{empty}


\section*{Introduction}

Over the past two decades, network-based approaches have emerged as a powerful framework for describing and analysing complex systems. By representing interactions among system components as graphs, this perspective has revealed universal organizing principles across domains ranging from technology and society to biology~\cite{Laci_revmod,Dorog_book,Newman_Barabasi_Watts,Jari_Holme_Phys_Rep,Vespignani_book}. Among these systems, neural networks, encoding the connections between neurons of living organisms, have always been of central interest, providing information on the structural organization and functional dynamics of the nervous system.
A widely known example is the neural network of the \textit{Caenorhabditis elegans} worm\cite{White_C_elegans,Cook_C_elegans_nature}, consisting of roughly 300-400 neurons (depending on the sex of the animal) with about 5,000-7,000 connections. A considerably larger network was reconstructed for the larva of the \textit{Drosophila}\cite{Winding_D_larva_brain_Science},  spanning between 3,016 neurons via roughly $5\cdot 10^5$ synapses. However, both of these networks are dwarfed by the recent reconstruction of the brain of adult female \textit{Drosophila melanogaster}\cite{FlyBrain_base_Nature,FlyBrain_NetStats_Nature}, which contains 139,255 neurons linked by $5\cdot 10^7$ chemical synapses. Given that flies are capable of navigating over distances\cite{Fly_navigate}, show signs of long-term memories\cite{Fly_long_memory}, engage in social interactions\cite{Fly_social}, and exhibit a wiring diagram between brain regions similar to that of mammals\cite{Fly_brain_similar_mammal_1,Fly_brain_similar_mammal_2}, research on the fly brain offers insights that extend beyond a mere increase in neural network scale.

The fundamental network characteristics of this fascinating system have already been examined\cite{FlyBrain_NetStats_Nature}, uncovering a scale-free degree distribution and a distinct rich-club organisation, in which highly central neurons (hubs) are densely interconnected. In addition, specific neuronal subsets were identified that may act as signal integrators or broadcasters. In the present study, we augment these results through the use of network embeddings, aimed at arranging the neurons in metric spaces solely based on the structure of the connections. In general, network embedding techniques provide an important alternative to traditional network measures for gaining information on various properties of the analysed network\cite{Goyal_Ferrara_embedding_survey,Radicchi_compare_embedding,Embedding_book_Yang,Bianconi_zoo_guide_to_embedding}. When transforming a network into a point cloud in a metric space, the original graph structure becomes encoded in the relative coordinates of the nodes, as, for example, tightly knit communities in the network are usually mapped onto compact and dense point clusters. 
The node coordinates also offer utility in several areas, including the prediction of missing links, assisting in navigation over the network, and serving as input for further machine learning tasks such as node classification, community finding, etc. 
Moreover, as demonstrated in Refs.~\cite{Kitsak2023HyperbolicMapping, zhihao_spa}, access to node coordinates can significantly aid in identifying nodes which contribute to shortest paths, especially in partially incomplete networks.

Although embedding nodes into the Euclidean space might seem as an intuitive choice, the hyperbolic approach provides a compelling alternative with distinct advantages\cite{Boguna_Krioukov_review}. Crucially, while Euclidean algorithms often require high-dimensional embeddings, hyperbolic approaches can achieve good quality embeddings in just two dimensions. This is because the exponential volume growth of hyperbolic spheres provides greater flexibility in node placement compared to the power-law growth of Euclidean spheres\cite{hyperGeomBasics}. The literature offers several different hyperbolic embedding algorithms, including likelihood optimization with respect to hyperbolic network models \cite{Boguna_Krioukov_Internet_2010,EPSO_HyperMap},  dimension reduction of non-linear Laplacian matrices \cite{Alanis-Lobato_LE_embedding,Alanis-Lobat_liekly_LE_emb} and Lorentz matrices using the hyperboloid model\cite{Hydra,our_dir_embedding}, coalescent embeddings\cite{coalescentEmbedding} (which apply dimension reduction to pre-weighted matrices capturing network structure), and mixed approaches combining dimension reduction and local optimization\cite{S1H2_Mercator,dmercator,our_embedding}, as well as neural network-based embeddings and approaches taking advantage of spanning trees~\cite{cannistraci_minimum_curvilinearity} or hierarchically nested communities in the network structure~\cite{commSector_hypEmbBasedOnComms_2019,CLOVE}. Most of these methods operate within the native representation of hyperbolic space, which in 2 dimensions, is often referred to as the native disk.

Nevertheless, most of the hyperbolic embedding methods above can not be scaled up to networks as large as the wiring diagram of the \textit{Drosophila melanogaster} brain~\cite{FlyBrain_base_Nature,FlyBrain_NetStats_Nature}, because of their substantial computational demands. As a result, a hyperbolic map of this system 
has remained unavailable so far. To overcome this limitation, we adopt the recently introduced  Cluster-Level Optimised Vertex Embedding (CLOVE) method~\cite{CLOVE}, which simultaneously delivers high embedding quality while maintaining exceptional computational efficiency. By avoiding the computational bottlenecks of alternative approaches, CLOVE makes it possible to construct a faithful hyperbolic embedding of the \textit{Drosophila melanogaster} brain connectome.
In parallel, we examine Euclidean embeddings produced by the similarly efficient Node2vec algorithm~\cite{node2vec}. We then systematically compare and evaluate the quality of the resulting representations using a range of metric scores. These embedding quality indicators are designed to quantify how well the spatial distribution of the nodes captures different aspects of the original network structure. The comparison between the quality of the embeddings of different geometries and different dimensions can give valuable information about the global organisation of the \textit{Drosophila} brain network. In addition to quantitative metrics, we utilize metadata available in the input to validate the embeddings. These labels categorize neurons into classes and super-classes (e.g., ''visual'', ''central'', etc.) that often exhibit strong spatial localization within the \textit{Drosophila} brain.

\section*{Results}

First, we embedded the \textit{Drosophila melanogaster} connectome \cite{FlyBrain_base_Nature,FlyBrain_NetStats_Nature} into the native disk representation of two-dimensional hyperbolic space using CLOVE~\cite{CLOVE}. According to Ref.~\cite{CLOVE}, a considerable advantage of CLOVE is its ability to provide high-quality hyperbolic embeddings in a very fast manner. A slight disadvantage, however, is that it currently operates only in two dimensions. To still uncover the role of the dimensionality of the underlying metric space, we also generated comparative Euclidean embeddings using Node2vec \cite{node2vec} in several dimensions ranging from 2 to 512. For more details about the applied embedding methods see the Methods section.

\begin{figure}[htbp]
    \centering

    \begin{subfigure}[t]{0.32\textwidth}
        \caption{\label{fig:subfigA}}
        \includegraphics[width=\textwidth]{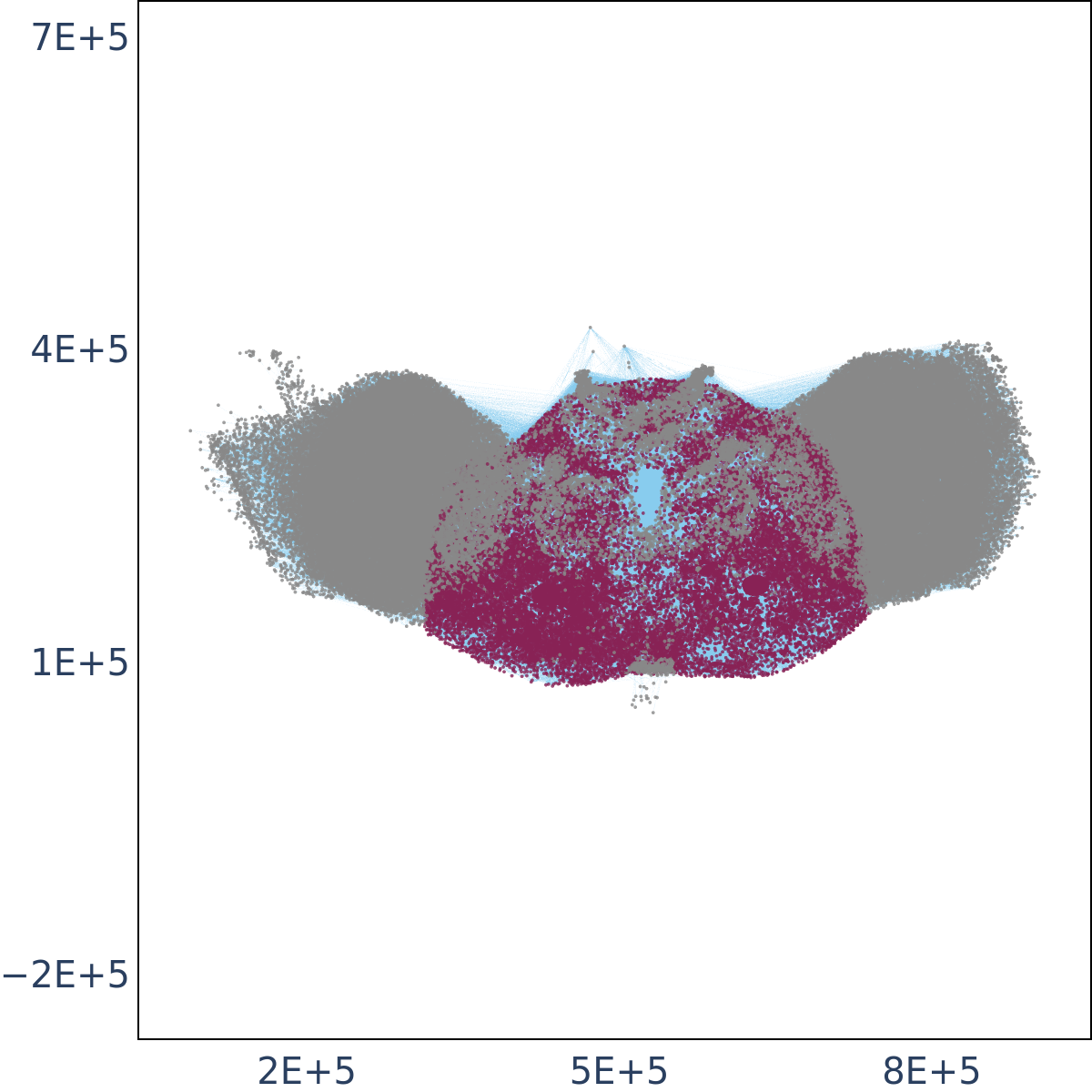}
    \end{subfigure}
    \begin{subfigure}[t]{0.32\textwidth}
        \caption{\label{fig:subfigB}}
        \includegraphics[width=\textwidth]{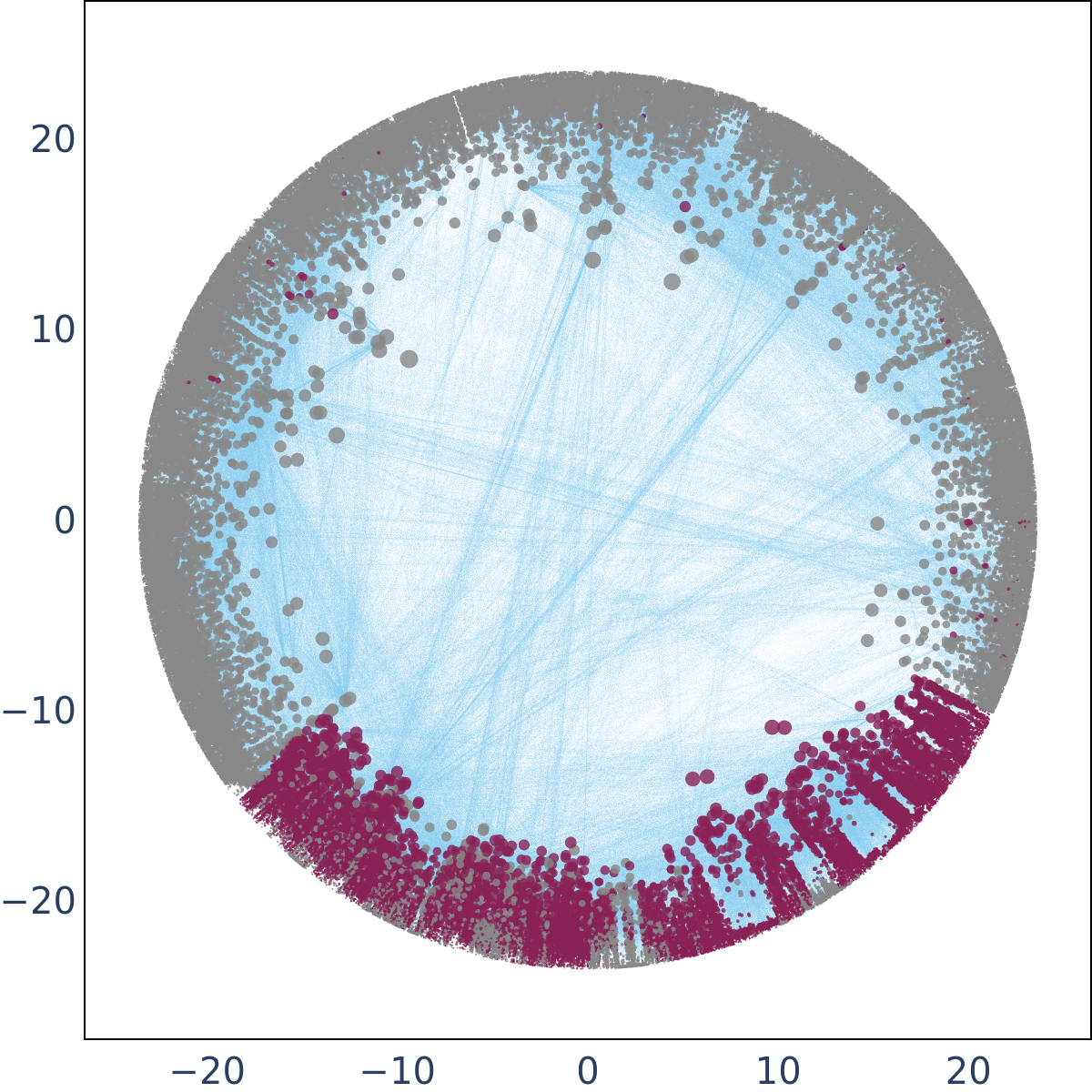}
    \end{subfigure}
    \begin{subfigure}[t]{0.32\textwidth}
        \caption{\label{fig:subfigC}}
        \includegraphics[width=\textwidth]{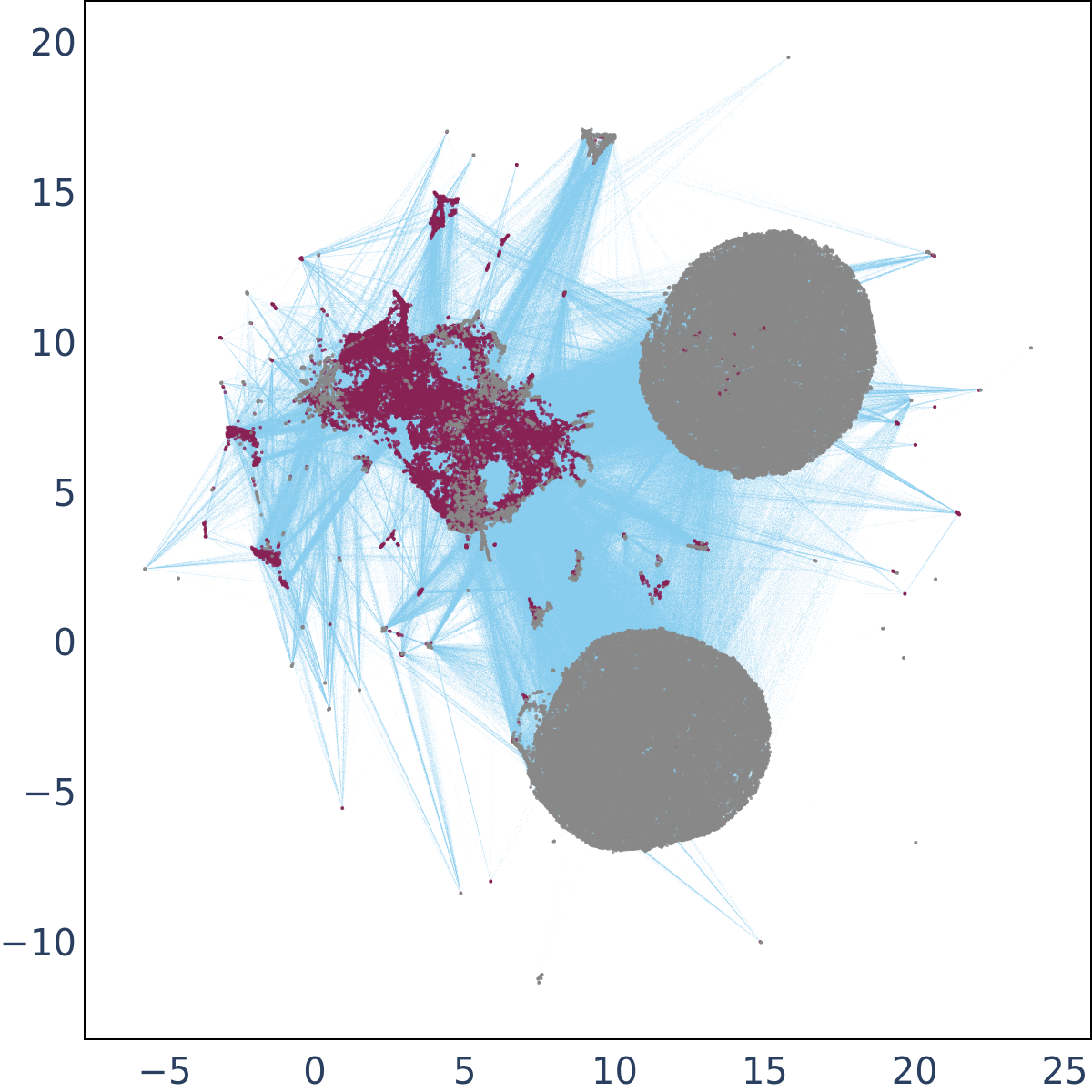}
    \end{subfigure}

    \vspace{1em}

    \begin{subfigure}[t]{0.32\textwidth}
        \caption{\label{fig:subfigD}}
        \includegraphics[width=\textwidth]{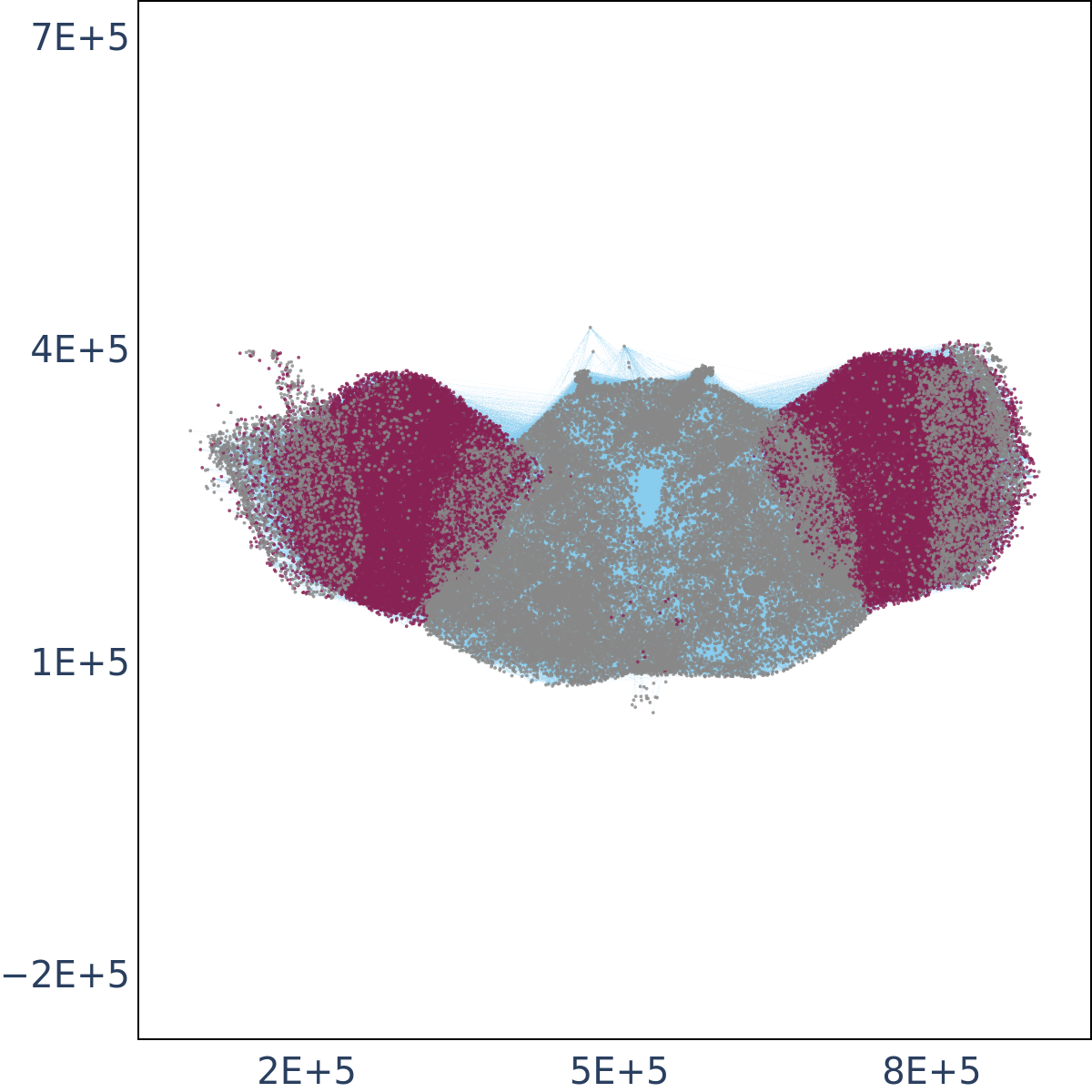}
    \end{subfigure}
    \begin{subfigure}[t]{0.32\textwidth}
        \caption{\label{fig:subfigE}}
        \includegraphics[width=\textwidth]{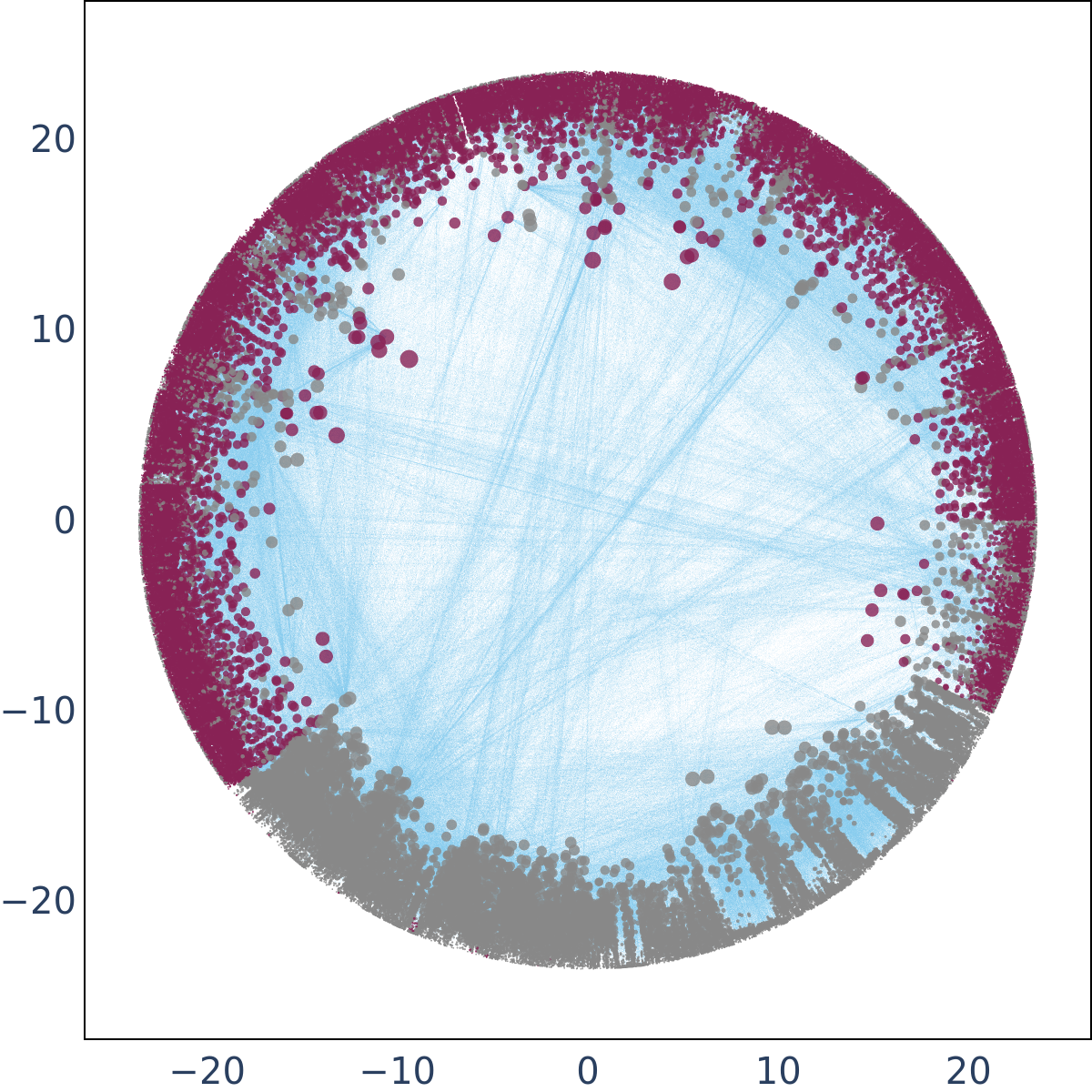}
    \end{subfigure}
    \begin{subfigure}[t]{0.32\textwidth}
        \caption{\label{fig:subfigF}}
        \includegraphics[width=\textwidth]{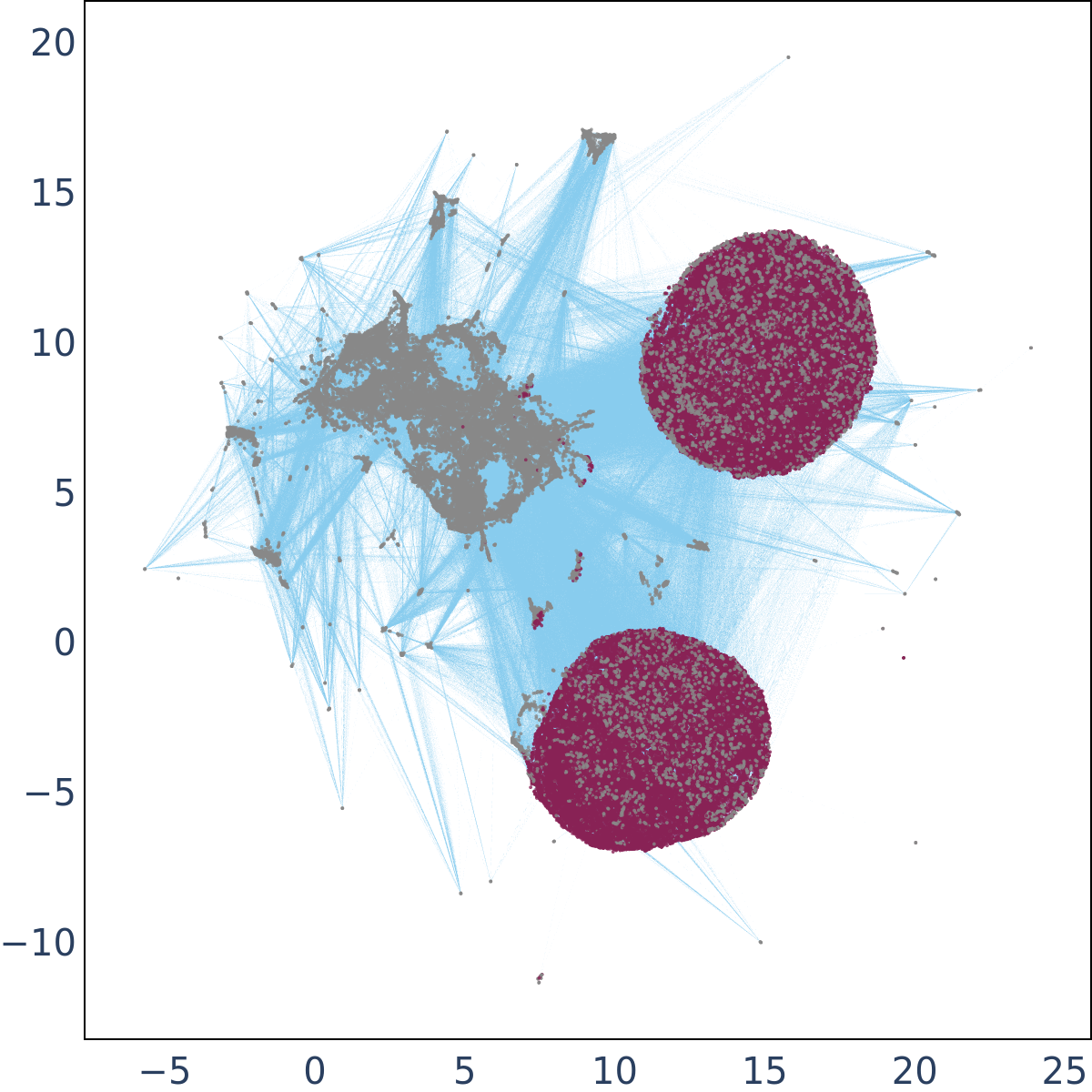}
    \end{subfigure}

    \caption{\textbf{Embeddings of the Drosophila brain network.} 
    In panels a) and d) (left column) we show a 2d Cartesian projection of the original neuron coordinates in the 3d Euclidean space.
    Panels b) and e) (middle column) display the hyperbolic embedding of the network in the 2d native disk representation according to CLOVE, where node size is an indicator of the degree. In panels c) and f) (right column) we display the 64d Euclidean embedding with Node2vec, projected to 2d using UMAP. In panels a), b), c) (top row) nodes highlighted in red correspond to neurons belonging to the "central" super class, whereas in panels d), e), f) (bottom row) the red node colour indicates that the given neuron belongs to the "optic" super class.}
    \label{fig:embedd_illustr}
\end{figure}

As an illustration of the obtained embeddings, in Fig.\ref{fig:embedd_illustr} we show the layout of the network both according to the original 3d Euclidean coordinates provided in the input data (left column, Fig.\ref{fig:embedd_illustr}a and Fig.\ref{fig:embedd_illustr}d), according to the 2d hyperbolic coordinates obtained with CLOVE (middle column, Fig.\ref{fig:embedd_illustr}b and Fig.\ref{fig:embedd_illustr}e) and a projection of the high dimensional Euclidean coordinates obtained with Node2vec (right column, Fig.\ref{fig:embedd_illustr}c and Fig.\ref{fig:embedd_illustr}f). In the latter case, we used the  Uniform Manifold Approximation and Projection\cite{umap} (UMAP) for projecting the 64d coordinates into 2 dimensions. Red node colour highlights the neurons belonging to the neuron super class labelled as "central"  in Fig.\ref{fig:embedd_illustr}a-c, whereas in Fig.\ref{fig:embedd_illustr}d-e we highlighted the super class "optic" for contrast.

\captionsetup[subfigure]{justification=raggedright,singlelinecheck=off}

Although Fig.\ref{fig:embedd_illustr}a and Fig.\ref{fig:embedd_illustr}d show only a 2d projection of the 3d structure of the Drosophila's brain, they still offer a good intuition of the system's large-scale organization, thanks to the dense placement of the nodes that clearly outline the silhouette of the brain. Moreover, we can also observe a relatively clean separation between the two neuron super classes highlighted. By comparison, the hyperbolic embedding (Fig.\ref{fig:embedd_illustr}b and Fig.\ref{fig:embedd_illustr}e), is much more spacious, allowing for the display of the links without compromising the visibility of the node layout. Here high degree nodes (often called as hubs) are placed closer to the centre of the native disk, while small degree nodes are pushed towards the disk periphery. We note that separation between the highlighted super classes we observed in the original Euclidean neuron coordinates is preserved by the angular arrangement of the nodes. 
Finally, according to the UMAP projections shown in Fig.\ref{fig:embedd_illustr}c and Fig.\ref{fig:embedd_illustr}f, the chosen super classes are also well separated in the 64d Euclidean embedding obtained with Node2vec. Such a clear separation of these larger-scale structural units for both embedding methods already supports the high quality of the inferred representations.

For a more detailed description of the superclasses, see the Methods section. The Supplementary Information S1 contains additional figures showing similar layouts with more neuronal superclasses highlighted.

\subsection*{Embedding quality measures}

To measure the quality of the obtained embeddings, we used well-established indicators from the scientific literature. These metrics assess the degree to which the resulting coordinates preserve the network's main structural features from different aspects. Among these, the Mapping Accuracy~\cite{Radicchi_compare_embedding}
, MA, quantifies the Spearman's rank correlation between the shortest path lengths and the pairwise geometric distance of the nodes. This measure is motivated by the expectation that node pairs close to each other in the network should be mapped to nearby points in the embedding space. 

It also follows intuitively from the property discussed above that the embedding coordinates can be highly informative for general tasks such as network reconstruction and missing link prediction, with the latter also commonly referred to as edge prediction~\cite{kitsak2020link, sinha2018systematic}.
In a simple framework, we may intend to reconstruct the network structure from the coordinates by starting from a completely empty graph and re-introduce connections one by one according to the order dictated by the node distances. The links actually existing in the original input network provide  positive ground truth classes, whereas unconnected node pairs in the input network correspond to negative ground truth classes for this procedure. Several quality indicators can be defined based on this framework, in the present study we used the the area under the  receiver operating characteristic curve
for edge prediction, EPAUC and the edge prediction precision, EPP, corresponding to the area under the precision recall curve during edge prediction. Furthermore, we also calculated two additional measures, corresponding to the fraction truly connected pairs between the top 20\% or top 5\% of the node pairs according to the shortest geometrical distance, denoted as Edge Prediction Recall  20\%, EPR20 and Edge Prediction Recall 5\%, EPR5, respectively.

Hyperbolic networks are known to facilitate effective navigation based on the node coordinates, where under ideal conditions, the shortest path between randomly chosen node pairs tends to lie close to the geodesic line connecting the  nodes in the embedding space. Greedy routing paths build upon this phenomenon by implementing a very simple routing protocol, where in the next step we always proceed to the neighbour who is the closest to the target node among all neighbours. This also means that greedy routing paths are not guaranteed to successfully reach the target node, i.e., they may instead run into a cycle of revisited neighbours where the greedy routing is stopped. A related embedding quality score is given by the Greedy Routing Success Rate~\cite{Boguna_2009_nat_phys}, GR, measuring the fraction of successful greedy routing paths that do eventually reach their targets. The Greedy Routing Score~\cite{coalescentEmbedding}, GRS, serving as another closely related metric also takes into account the lengths of the paths, where the ratio between the geometrical length of the topological shortest path and that of the greedy path is averaged over all node pairs. Finally, the Greedy Routing Efficiency~\cite{Carlo_Nat_coms_hyp_congruency}, GRE metric evaluates the relationship between geodesic distances and projected greedy paths. For a more detailed description of the metric scores above, see the Methods section.

In Table \ref{table:low_dim_embeddings} we show the measured embedding quality indicators for low dimensional embeddings, including both the original real 3D neuron coordinates, the 2d hyperbolic embedding according to CLOVE and the Euclidean network embeddings obtained with Node2vec in dimensions $d=2$ and $d=3$. We note that due to the large size of the studied network, exact evaluation of the quality measures is not feasible under reasonable time. Accordingly, our results are instead based on random sampling, a convenient method that was shown to provide a reasonably accurate estimate of the quality measures in a previous work\cite{CLOVE}. 


\begin{table}[hbt!]
\begin{tabular}{?l?c|c|c|c|c|c|c|c?}
		\thickhline
		& \makecell{MA} & \makecell{GR} & \makecell{GRS} & \makecell{GRE} & \makecell{EPAUC} & \makecell{EPP} & \makecell{EPR20} & \makecell{EPR5}\\
		\thickhline
		\hline
		\makecell{Drosophila brain \\real 3d} & \cellcolor[RGB]{192.000, 230.000, 185.000}$0.363$ & \cellcolor[RGB]{239.000, 249.000, 236.000}$0.075$ & \cellcolor[RGB]{242.000, 250.000, 239.000}$0.048$ & \cellcolor[RGB]{242.000, 250.000, 239.000}$0.050$ & \cellcolor[RGB]{60.000, 165.000, 89.000}$0.862$ & \cellcolor[RGB]{62.000, 168.000, 91.000}$0.849$ & \cellcolor[RGB]{52.000, 157.000, 83.000}$0.904$ & \cellcolor[RGB]{41.000, 145.000, 74.000}$0.968$ \\
		\hline
		\makecell{Node2vec 2d \\(Euclidean)} & \cellcolor[RGB]{192.000, 230.000, 185.000}$0.364$ & \cellcolor[RGB]{244.000, 251.000, 241.000}$0.030$ & \cellcolor[RGB]{244.000, 251.000, 242.000}$0.023$ & \cellcolor[RGB]{244.000, 251.000, 242.000}$0.026$ & \cellcolor[RGB]{61.000, 167.000, 90.000}$0.853$ & \cellcolor[RGB]{72.000, 175.000, 97.000}$0.809$ & \cellcolor[RGB]{65.000, 171.000, 93.000}$0.832$ & \cellcolor[RGB]{53.000, 159.000, 84.000}$0.898$ \\
		\hline
		\makecell{Node2vec 3d \\(Euclidean)} & \cellcolor[RGB]{166.000, 219.000, 160.000}$0.476$ & \cellcolor[RGB]{240.000, 249.000, 237.000}$0.061$ & \cellcolor[RGB]{242.000, 250.000, 239.000}$0.047$ & \cellcolor[RGB]{242.000, 250.000, 239.000}$0.046$ & \cellcolor[RGB]{49.000, 154.000, 81.000}$0.920$ & \cellcolor[RGB]{53.000, 158.000, 83.000}$0.902$ & \cellcolor[RGB]{47.000, 152.000, 78.000}$0.934$ & \cellcolor[RGB]{42.000, 147.000, 75.000}$0.959$ \\
		\hline
		\makecell{CLOVE \\(hyperbolic)} & \cellcolor[RGB]{154.000, 214.000, 149.000}$\mathbf{0.528}$ & \cellcolor[RGB]{147.000, 210.000, 143.000}$\mathbf{0.553}$ & \cellcolor[RGB]{186.000, 228.000, 179.000}$\mathbf{0.390}$ & \cellcolor[RGB]{230.000, 245.000, 225.000}$\mathbf{0.160}$ & \cellcolor[RGB]{42.000, 147.000, 75.000}$\mathbf{0.960}$ & \cellcolor[RGB]{41.000, 146.000, 74.000}$\mathbf{0.964}$ & \cellcolor[RGB]{36.000, 140.000, 70.000}$\mathbf{0.996}$ & \cellcolor[RGB]{35.000, 139.000, 69.000}$\mathbf{1.000}$ \\
		\thickhline
\end{tabular}
	\caption{ \textbf{Embedding quality scores for the low dimensional embeddings.} We show the results obtained via sampling 0.001 fraction of the node pairs at random for the mapping accuracy, MA, the edge Greedy Routing Succes Rate, GR, the Greedy Routing Score, GRS, the Greedy Routing Efficiency, GRE, the Edge Prediction Area Under the Receiver Operating Characteristic Curve, EPAUC, the Edge Prediction Precision, EPP, as well as the EPR20 and the EPR5 scores. The measured quality indicators are presented for the original 3d neuron positions in the Drosophila brain, the Euclidean network embeddings obtained with Node2vec in 3 and in 2 dimensions and the 2d hyperbolic embedding obtained with Cluster Level Optimized Vertex Embedding (CLOVE).}
	\label{table:low_dim_embeddings}
\end{table}

Table \ref{table:low_dim_embeddings} indicates that the 2d hyperbolic embedding obtained with CLOVE achieves higher scores than the low-dimensional Euclidean embeddings across all evaluated quality metrics. Moreover, all scores for the 2d hyperbolic embedding consistently surpass those of the real physical 3D arrangement. In contrast, the two-dimensional Node2vec embedding never reaches this level of performance, and even the three-dimensional Node2vec embedding performs worse than the real physical 3D configuration for several of the considered metrics.

The difference between the quality of the hyperbolic embedding and the Euclidean counterparts is particularly large in the case of the scores related to greedy navigation on the network (GR, GRS and GRE). This clearly shows that the hyperbolic map of the \textit{Drosophila melanogaster} connectome is substantially more navigable than low-dimensional Euclidean representations, thereby enabling efficient greedy information transfer along geodesic curves in the underlying metric space.

To uncover the role of dimensionality, In Table \ref{table:high_dim_nod2vec} we display the same quality scores for Euclidean Node2vec embeddings where the dimension is varying between $d=4$ and $d=512$.
\begin{table}[hbt!]
	\begin{tabular}{?l?c|c|c|c|c|c|c|c?}
		\thickhline
		& \makecell{MA} & \makecell{GR} & \makecell{GRS} & \makecell{GRE} & \makecell{EPAUC} & \makecell{EPP} & \makecell{EPR20} & \makecell{EPR5}\\
		\thickhline
		\hline
		\makecell{Node2vec 4d (Euclidean)} & \cellcolor[RGB]{161.000, 217.000, 155.000}$0.500$ & \cellcolor[RGB]{232.000, 246.000, 227.000}$0.142$ & \cellcolor[RGB]{236.000, 248.000, 232.000}$0.102$ & \cellcolor[RGB]{236.000, 248.000, 233.000}$0.098$ & \cellcolor[RGB]{45.000, 149.000, 77.000}$0.947$ & \cellcolor[RGB]{46.000, 151.000, 78.000}$0.938$ & \cellcolor[RGB]{41.000, 145.000, 74.000}$0.967$ & \cellcolor[RGB]{38.000, 143.000, 72.000}$0.982$ \\
		\hline
		\makecell{Node2vec 8d (Euclidean)} & \cellcolor[RGB]{130.000, 203.000, 130.000}$0.613$ & \cellcolor[RGB]{175.000, 223.000, 169.000}$0.438$ & \cellcolor[RGB]{201.000, 234.000, 194.000}$0.322$ & \cellcolor[RGB]{213.000, 239.000, 207.000}$0.256$ & \cellcolor[RGB]{38.000, 142.000, 71.000}$0.986$ & \cellcolor[RGB]{38.000, 143.000, 72.000}$0.982$ & \cellcolor[RGB]{37.000, 141.000, 70.000}$0.991$ & \cellcolor[RGB]{36.000, 140.000, 70.000}$0.993$ \\
		\hline
		\makecell{Node2vec 16d (Euclidean)} & \cellcolor[RGB]{120.000, 198.000, 121.000}$0.653$ & \cellcolor[RGB]{126.000, 201.000, 126.000}$0.629$ & \cellcolor[RGB]{165.000, 219.000, 159.000}$0.484$ & \cellcolor[RGB]{201.000, 234.000, 194.000}$0.323$ & \cellcolor[RGB]{36.000, 140.000, 70.000}$0.993$ & \cellcolor[RGB]{37.000, 141.000, 70.000}$0.990$ & \cellcolor[RGB]{36.000, 140.000, 70.000}$0.995$ & \cellcolor[RGB]{36.000, 140.000, 69.000}$0.997$ \\
		\hline
		\makecell{Node2vec 32d (Euclidean)} & \cellcolor[RGB]{118.000, 197.000, 119.000}$\mathbf{0.660}$ & \cellcolor[RGB]{85.000, 181.000, 103.000}$0.767$ & \cellcolor[RGB]{131.000, 203.000, 131.000}$0.609$ & \cellcolor[RGB]{195.000, 231.000, 188.000}$\mathbf{0.352}$ & \cellcolor[RGB]{36.000, 140.000, 70.000}$0.995$ & \cellcolor[RGB]{36.000, 140.000, 70.000}$0.994$ & \cellcolor[RGB]{36.000, 140.000, 69.000}$0.997$ & \cellcolor[RGB]{35.000, 139.000, 69.000}$0.998$ \\
		\hline
		\makecell{Node2vec 64d (Euclidean)} & \cellcolor[RGB]{123.000, 199.000, 123.000}$0.642$ & \cellcolor[RGB]{63.000, 168.000, 91.000}$0.847$ & \cellcolor[RGB]{110.000, 193.000, 115.000}$0.687$ & \cellcolor[RGB]{195.000, 231.000, 188.000}$0.349$ & \cellcolor[RGB]{36.000, 140.000, 70.000}$\mathbf{0.996}$ & \cellcolor[RGB]{36.000, 140.000, 70.000}$\mathbf{0.995}$ & \cellcolor[RGB]{35.000, 139.000, 69.000}$\mathbf{0.998}$ & \cellcolor[RGB]{35.000, 139.000, 69.000}$\mathbf{0.999}$ \\
		\hline
		\makecell{Node2vec 128d (Euclidean)} & \cellcolor[RGB]{146.000, 210.000, 143.000}$0.554$ & \cellcolor[RGB]{60.000, 166.000, 89.000}$\mathbf{0.859}$ & \cellcolor[RGB]{103.000, 190.000, 112.000}$\mathbf{0.709}$ & \cellcolor[RGB]{200.000, 233.000, 193.000}$0.327$ & \cellcolor[RGB]{36.000, 140.000, 70.000}$0.996$ & \cellcolor[RGB]{36.000, 140.000, 70.000}$\mathbf{0.995}$ & \cellcolor[RGB]{35.000, 139.000, 69.000}$\mathbf{0.998}$ & \cellcolor[RGB]{35.000, 139.000, 69.000}$\mathbf{0.999}$ \\
		\hline
		\makecell{Node2vec 256d (Euclidean)} & \cellcolor[RGB]{179.000, 225.000, 172.000}$0.422$ & \cellcolor[RGB]{88.000, 182.000, 104.000}$0.759$ & \cellcolor[RGB]{126.000, 201.000, 126.000}$0.629$ & \cellcolor[RGB]{211.000, 238.000, 204.000}$0.269$ & \cellcolor[RGB]{36.000, 140.000, 70.000}$0.994$ & \cellcolor[RGB]{36.000, 140.000, 70.000}$0.992$ & \cellcolor[RGB]{36.000, 140.000, 70.000}$0.996$ & \cellcolor[RGB]{36.000, 140.000, 69.000}$0.997$ \\
		\hline
		\makecell{Node2vec 512d (Euclidean)} & \cellcolor[RGB]{229.000, 245.000, 224.000}$0.165$ & \cellcolor[RGB]{163.000, 218.000, 157.000}$0.493$ & \cellcolor[RGB]{182.000, 226.000, 176.000}$0.407$ & \cellcolor[RGB]{229.000, 245.000, 224.000}$0.169$ & \cellcolor[RGB]{37.000, 141.000, 71.000}$0.987$ & \cellcolor[RGB]{38.000, 143.000, 72.000}$0.981$ & \cellcolor[RGB]{37.000, 141.000, 71.000}$0.988$ & \cellcolor[RGB]{37.000, 141.000, 71.000}$0.987$ \\
		\thickhline
	\end{tabular}
	\caption{\textbf{Embedding quality scores for high dimensional Node2vec embeddings.} We display the measured average quality scores for he Mapping Accuracy, MA, the edge Greedy Routing Succes Rate, GR, the Greedy Routing Score, GRS, the Greedy Routing Efficiency, GRE, the Edge Prediction Area Under the Receiver Operating Characteristic Curve, EPAUC, the Edge Prediction Precision, EPP, and finally, the EPR20 and the EPR5 scores. The rows correspond to the different embedding dimensions as indicated in the leftmost column from $d=4$ to $d=512$.}
	\label{table:high_dim_nod2vec}
\end{table}
The results indicate that quality scores start rapidly increasing with the embedding dimension, then eventually reach a peak, and then transit into decline for very high dimensional embeddings. When compared with the scores achieved by the 2d hyperbolic embedding, for most quality indicators the level set by the CLOVE result is surpassed somewhere between $d=8$ and $d=16$. However, for link prediction -- measured by metrics such as EPR20 and EPR5 -- CLOVE maintains competitive performance even against very high-dimensional Node2vec embeddings. For these scores, both the 2d hyperbolic embedding and the high dimensional Euclidean Node2vec embeddings (in the range $d=16$ - $d=256$) closely approach 1.

\begin{figure}[htbp]
    \centering
    \begin{subfigure}[t]{0.45\textwidth}
        \caption{\label{fig:subfigBa}}
        \includegraphics[width=\textwidth]{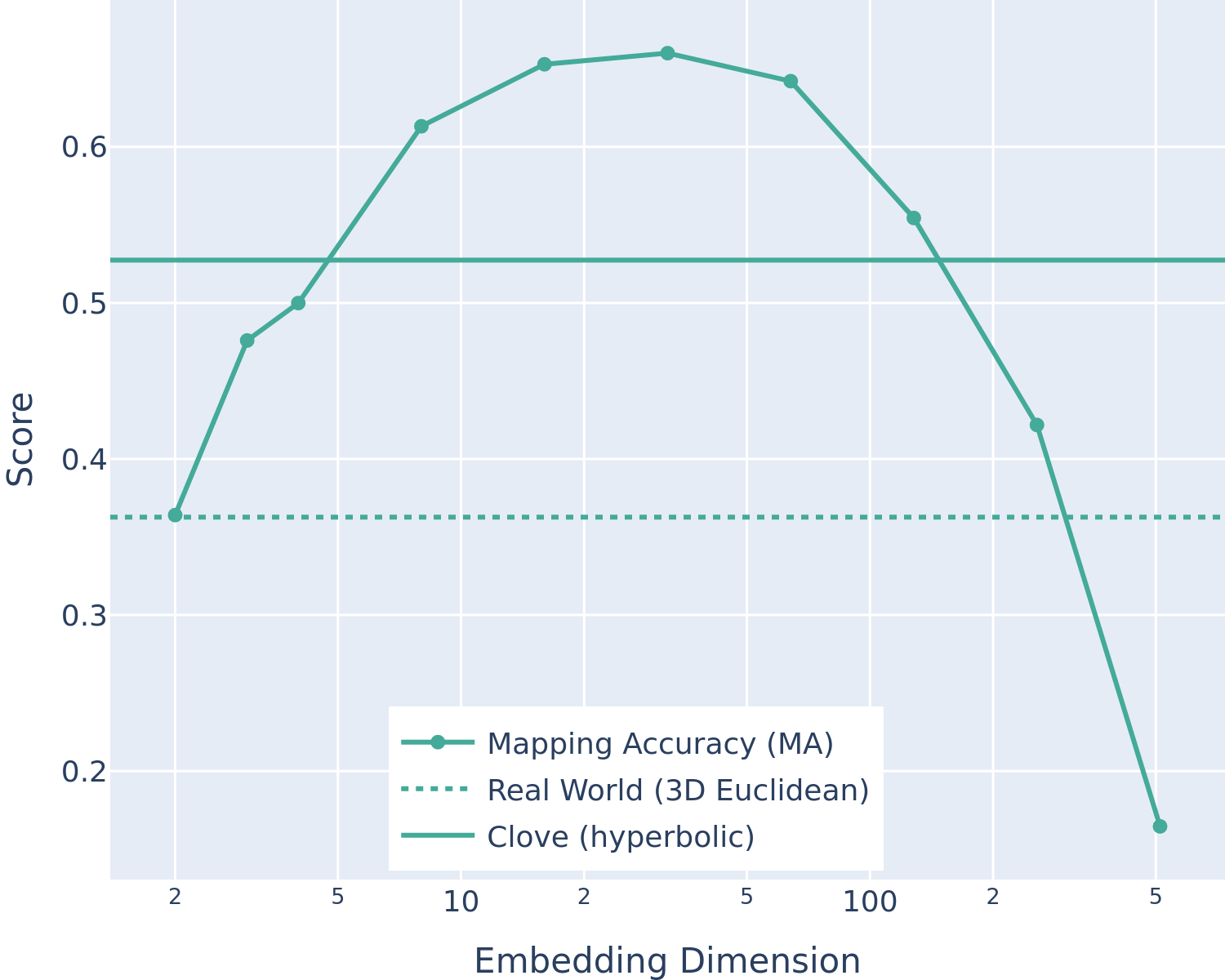}
    \end{subfigure}
    \begin{subfigure}[t]{0.45\textwidth}
        \caption{\label{fig:subfigBb}}
        \includegraphics[width=\textwidth]{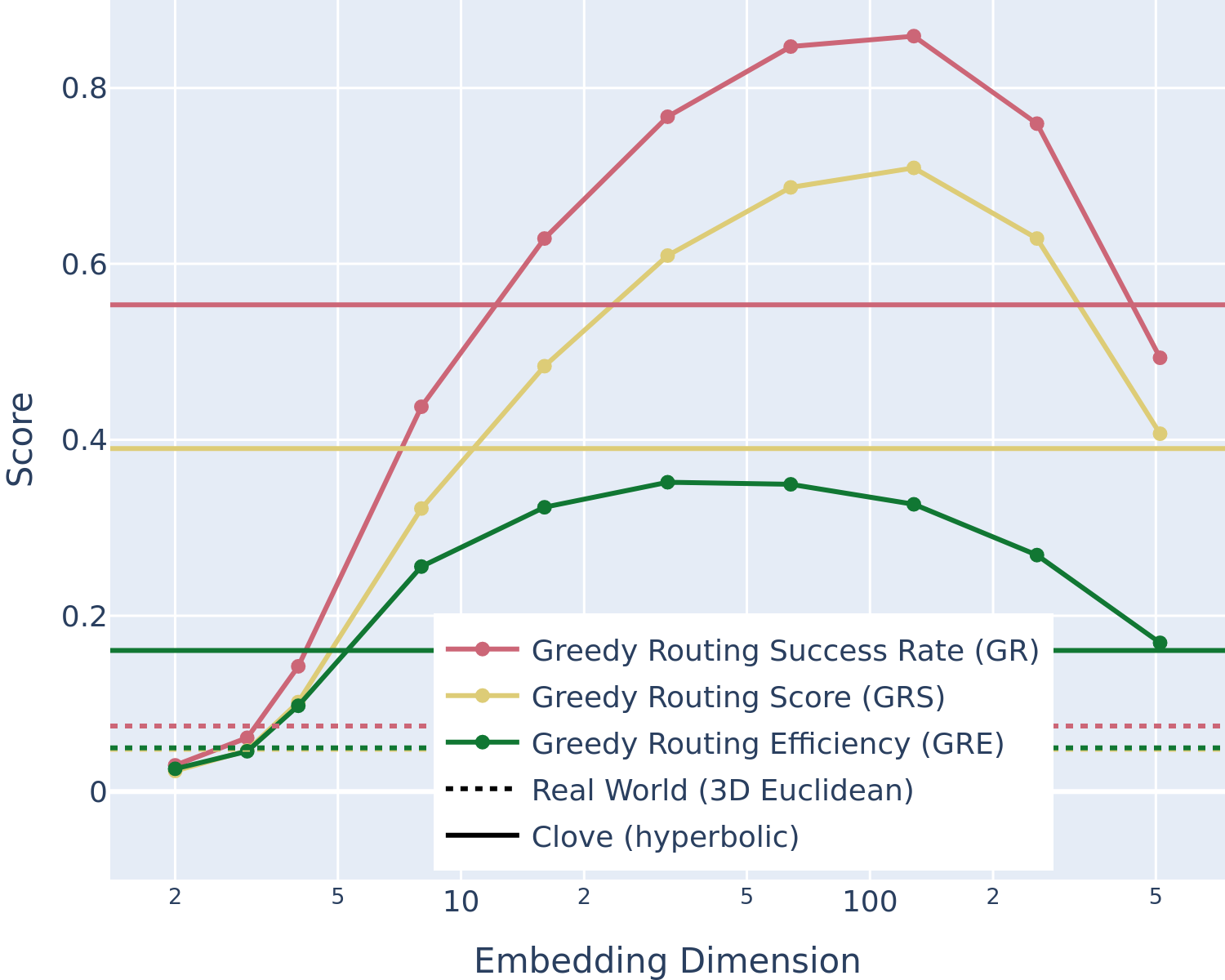}
    \end{subfigure}
    
    \begin{subfigure}[t]{0.45\textwidth}
        \caption{\label{fig:subfigBc}}
        \includegraphics[width=\textwidth]{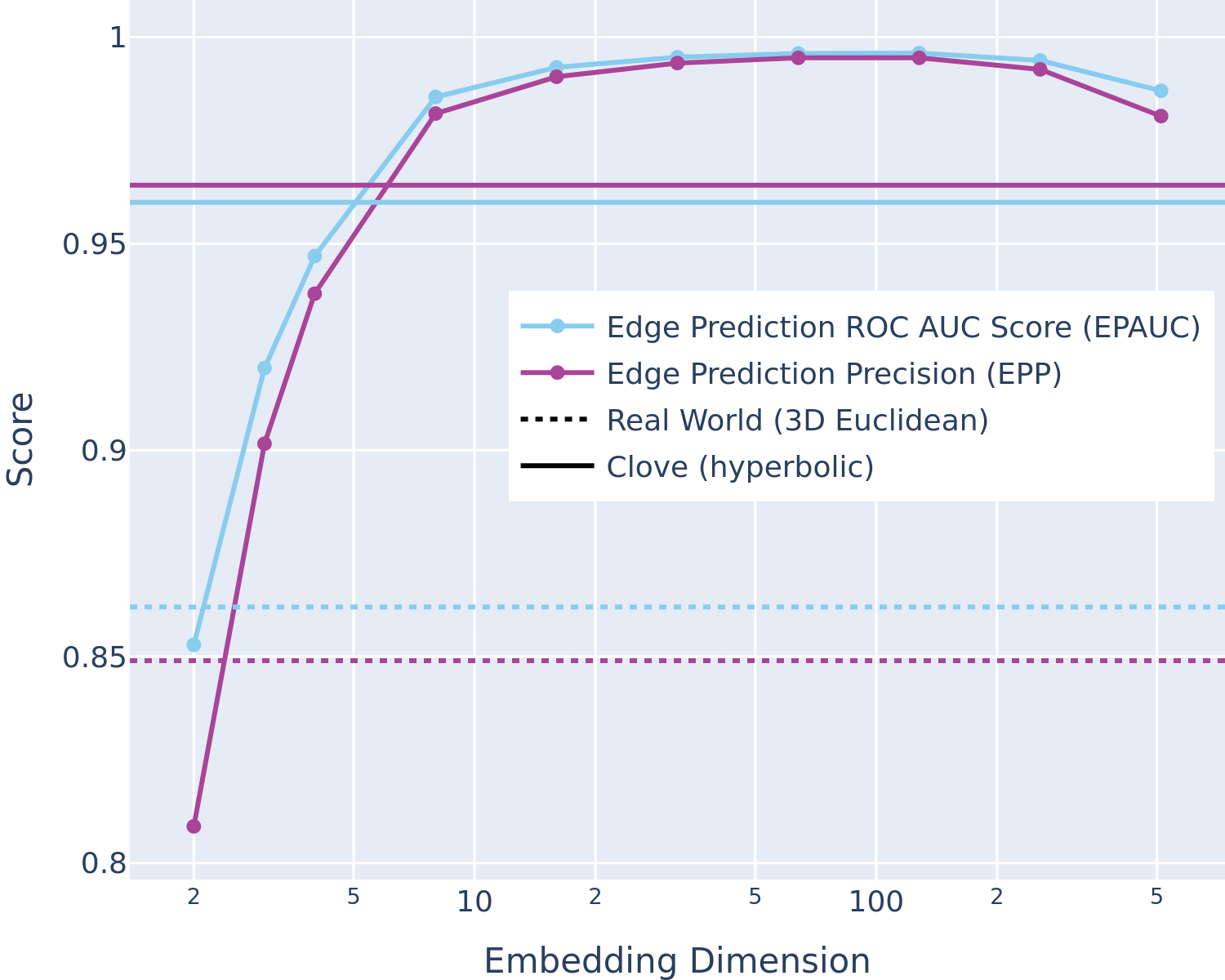}
    \end{subfigure}
    \begin{subfigure}[t]{0.45\textwidth}
        \caption{\label{fig:subfigBd}}
        \includegraphics[width=\textwidth]{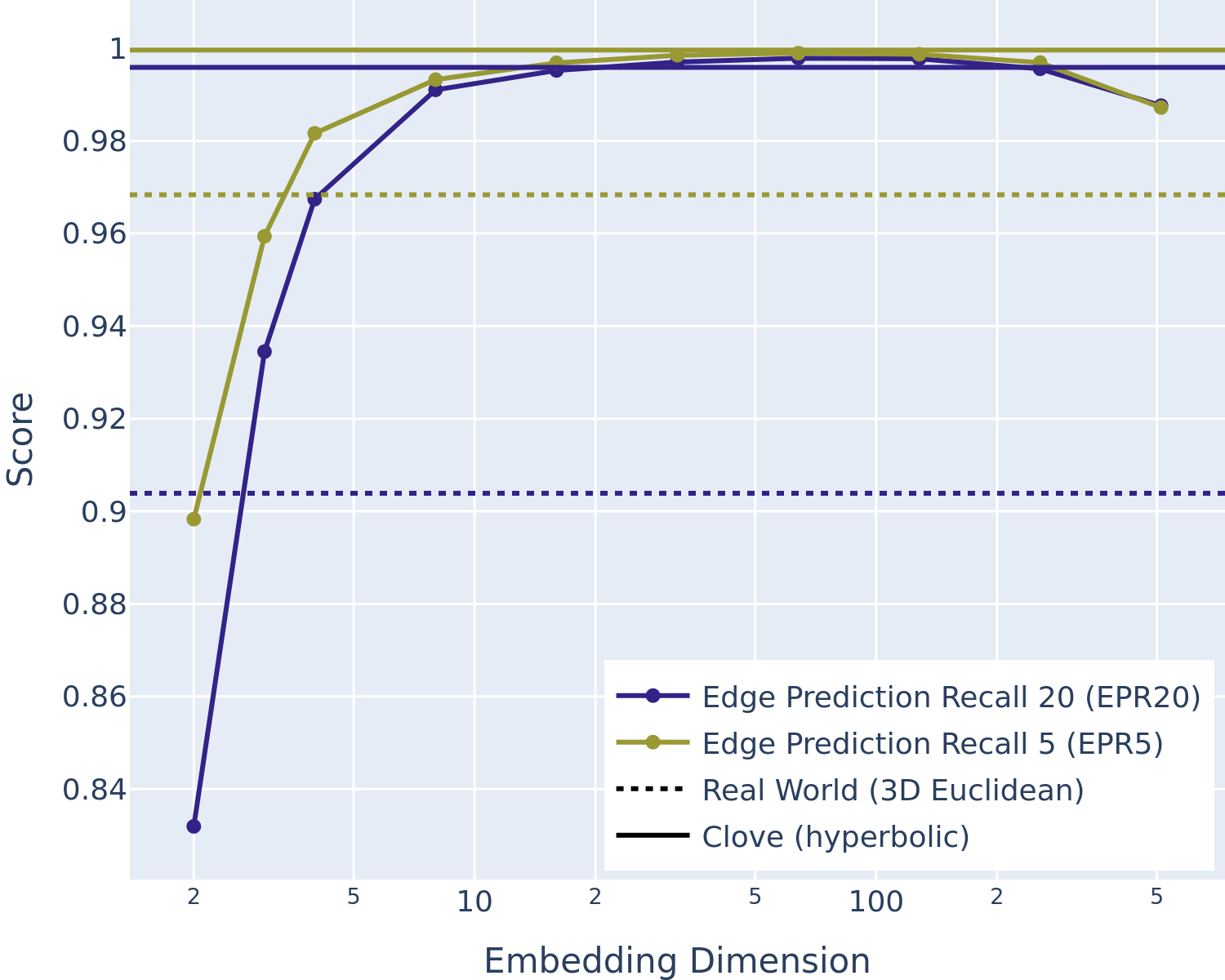}
    \end{subfigure}
    \caption{\textbf{Embedding quality scores compared across embedding dimensions.} We show the scores obtained for Node2vec as a function of the embedding dimension with line plots, compared with the result for the 2d hyperbolic embedding (solid horizontal line) and for the original 3d neuron coordinates (dashed horizontal line). a) The Mapping Accuracy. b) The scores related to greedy navigation: the Greedy Routing Succes Rate (light blue) the Greedy Routing Score (burgundy) and the Greedy Routing Efficiency (light green). c) The Area Under the Receiver Operating Characterstic Curve for edge prediction, EPAUC (light blue) and the Edge Prediction Precision (burgundy). d) The Edge Prediction Recall results for EPR20 (dark green) and EPR5 (light green).}
    \label{fig:score_vs_dim}
\end{figure}

To provide a view on these results from an additional perspective, in Fig.\ref{fig:score_vs_dim} we plotted the embedding quality scores as a function of the embedding dimension for the Node2vec embeddings. Notably, this figure clearly confirms that all types of embedding quality indicator, without an exception, exhibit a non-monotonic behaviour as a function of embedding dimension. Nevertheless, the optimal embedding dimension is not universal and varies depending on the specific evaluation metric. In addition, Fig.~\ref{fig:score_vs_dim} shows that the Mapping Accuracy (Fig.\ref{fig:score_vs_dim}a) and the scores related to the greedy navigation (Fig.\ref{fig:score_vs_dim}b) have a more pronounced maximum compared to the scores related to missing link prediction (Fig.\ref{fig:score_vs_dim}c-d), which seem to display a relatively wide plateau between $d=16$ and $d=256$. Based on these plots and Table \ref{table:high_dim_nod2vec}, the optimal choice for the dimension in the case of Node2vec where most of the quality scores are close to their best is either $d=64$ or $d=128$. 

\captionsetup[subfigure]{justification=raggedright,singlelinecheck=off}

\section*{Discussion}

Investigating the organisational principles and functional dynamics of biological neural networks is central to unravelling the fundamental mechanisms of information processing and adaptive behaviour. 
Prior studies have suggested that the brain’s wiring can exhibit various signatures consistent with hyperbolic geometry~\cite{joseph2024hyperbolic}. Analyses of the human brain’s functional architecture also reveal a hierarchically modular structure~\cite{Lambiotte_hier_modul_brain,Mastrandrea2017_hier_brain}, a feature that naturally emerges in hyperbolic networks~\cite{our_hyp_coms,our_modular_pso,our_ultra_cold_modular_pso}. Brain regions coordinate through a highly hierarchical, chain-like arrangement of clustered anatomical zones. Individual neurons, with their branching axons and dendrites, resemble trees that collectively form a vast neural forest~\cite{ramonycajal1995histology,wen2005segregation,cuntz2010onerule}. Given that the hyperbolic space is fundamentally a continuous representation of tree structures, it serves as an ideal geometric model for this network.

The analysis of the navigability of the map of neural connections in the brain across various species showed a striking difference between the navigability properties of mammalian and non-mammalian species, which implies the inability of Euclidean distances to fully explain the structural organization of their neural connectomes~\cite{Allard2020_brain_net_navig}. In contrast, hyperbolic space provided almost perfectly navigable maps for these connectomes for all species, showing that hyperbolic distances are exceptionally congruent with the structure of brain networks~\cite{Allard2020_brain_net_navig}. The significance of hyperbolic spaces was also highlighted in neural signalling networks, where this geometric framework offered enhanced robustness~\cite{Sharpee2019_hyp_neural_circuits}. 

The latent geometry of neural networks is often studied with the help of node embedding techniques that project the nodes into a geometric space based on the network structure. The human brain connectome -- at both the structural and functional levels -- has been successfully embedded in into hyperbolic spaces~\cite{Carlo_2017_coalescent,Whi_2022_hyp_embed_brain}. Following a similar line, in the present work we examined the geometry of the \textit{Drosophila melanogaster} connectome using both Euclidean and hyperbolic embeddings. We observed that major neuronal superclasses, which are spatially distinct in the fly’s 3D anatomy, map to separate regions within the geometric embedding. These findings, supported by high embedding quality scores, demonstrate that these methods generate meaningful representations where the spatial distribution of nodes effectively encodes the brain's structural organization.

A detailed analysis of embedding quality scores revealed that the 2d hyperbolic embedding (obtained via CLOVE\cite{CLOVE}) outperforms both the original physical 3D arrangement of neurons and the 2d or 3d Euclidean embeddings (obtained with Node2vec\cite{node2vec}) across all metrics. When examining higher-dimensional Euclidean embeddings, quality scores increase rapidly as a function of dimension, surpassing the hyperbolic baseline at approximately $d=8$. These scores peak between $d=32$ and $d=128$, depending on the metric, before starting to decline in very high-dimensional regimes. While the absolute highest scores were achieved with medium-to-high dimensional Euclidean embeddings, it is important to note that comparing embeddings across vastly different dimensions is nuanced; the additional degrees of freedom in higher dimensions provide an inherent representational advantage over low-dimensional spaces.

In summary, our analysis showcases that the organization of neural connectivity in the \textit{Drosophila melanogaster} brain is significantly more congruent with hyperbolic geometry than Euclidean geometry in low-dimensional (2d or 3d) spaces. These findings corroborate previous research on the hyperbolic nature of neural networks, reinforcing the view that this geometry offers a natural framework for capturing their hierarchical and modular organization.

\section*{Methods and data}

\subsection*{Embedding quality scores}

In this section, we provide a detailed description of the embedding quality metrics used to evaluate how accurately the network is represented within the underlying metric space.

\subsubsection*{Greedy Routing based metrics}

Greedy routing \cite{milgram1967small, kleinberg2000small,Boguna_2009_nat_phys} is a myopic path-finding strategy in which a packet is sent from a source node to a target node, making decisions based solely on local information. At each step, the packet is only aware of the target's coordinates and the neighbours of the current node in the embedding space, and it is forwarded to the neighbour closest to the target. The efficiency of this routing procedure is evaluated across all possible source–target pairs, or a representative subset thereof. We consider three variants of this metric with different levels of granularity. Although computationally intensive, these measures provide a robust and informative assessment of how well an embedding facilitates navigation and efficient path finding~\cite{sulyok2023greedy}.

The first metric, the Greedy Routing success rate (GR), measures the fraction of packets that successfully reach their target~\cite{Boguna_2009_nat_phys}:
\begin{equation}
    \text{GR}
    =
    \frac{1}{N (N-1)}
    \sum_{s \in V} \sum_{\substack{t \in V \\ t \neq s}} \delta_{s \rightarrow t},
    \label{eq:GSDef}
\end{equation}
where $\delta_{s \rightarrow t} = 1$ if the greedy routing from sources $s$ to target $t$ is successful, otherwise $\delta_{s \rightarrow t} = 0$.

The second metric, the Greedy Routing score (GRS), introduces an additional level of granularity by accounting not only for whether a path is successful, but also for its length~\cite{coalescentEmbedding}. Specifically, successful paths are weighted by the ratio between the shortest-path length (in the underlying topology) connecting the source and target nodes and the actual path length produced by the greedy routing process. More precisely, it is defined as~\cite{coalescentEmbedding}:
\begin{equation}
{\rm GRS} = \frac{1}{N(N-1)}\cdot\sum\limits_{s\in N}\,\,\sum\limits_{t\in N, t\neq s}\frac{\ell_{s\rightarrow t}^{\mathrm{TSP}}}{\ell_{s\rightarrow t}^{\mathrm{GR}}}, 
\label{eq:GRscoreDef}
\end{equation}
with $\ell_{s\rightarrow t}^{\mathrm{TSP}}$ and $\ell_{s\rightarrow t}^{\mathrm{GR}}$ denoting the topological length of shortest path and greedy path respectively.

Finally, the Greedy Routing Efficiency (GRE) metric places even greater emphasis on geometric efficiency, rewarding not only successful deliveries but also paths that closely approximate the shortest possible routes in the network~\cite{Carlo_Nat_coms_hyp_congruency}. Successful paths are now weighted with fractions of the metric distance between the source and target nodes and the sum of distances travelled in the metric space during the greedy routing protocol~\cite{Carlo_Nat_coms_hyp_congruency}:
\begin{equation}
{\rm GR} = \frac{1}{N(N-1)}\cdot\sum\limits_{s\in N}\,\,\sum\limits_{t\in N, t\neq s}\frac{\Delta_{s\rightarrow t}^{GEOM}}{\Delta_{s\rightarrow t}^{GR}},
\label{eq:GRscoreDef2}
\end{equation}
where $\Delta_{s\rightarrow t}^{\mathrm{GEOM}}$ denotes the geodesic distance between the source node $s$ and the target node $t$ in the embedding space, while $\Delta_{s\rightarrow t}^{\mathrm{GR}}$ represents the total geometric length of the greedy path, computed as the sum of the edge lengths along the route followed by the greedy routing protocol.

\subsubsection*{Mapping Accuracy}
Mapping accuracy \cite{Radicchi_compare_embedding} is defined as the Spearman rank correlation between the topological shortest-path distances and the corresponding geodesic distances in the embedding space:
\begin{equation}
{\rm MA} = \rho \Bigl[ R \left[ TSP \right], R \left[ GEOM \right] \Bigr], 
\label{eq:MADef}
\end{equation}
where $R$ denotes the rank operator, $\rho$ the standard Pearson correlation, $TSP$ and $GEOM$ are arrays containing the topological shortest path and geometrical distances of node pairs, respectively.

\subsubsection*{Edge prediction}

An important property of a high quality embedding is that it assigns direct neighbours close to each other in the underlying space~\cite{kitsak2020link, sinha2018systematic}. To evaluate the quality of the learned embeddings from this perspective, we construct a unified evaluation set $S$ consisting of a set of positive edges $E^+$ (actual links) and a set of negative edges $E^-$ (non-existent links). For every edge $e_{ij} \in S$ connecting nodes $i$ and $j$, we compute a similarity score $s_{ij}$ based on their distance in the embedding space. The elements of the set $S$ are then ranked in descending order of their $s_{ij}$ values.

In \textit{EPAUC} we compute the Area Under the ROC Curve -- a popular method of assessing the predictive performance in different machine learning tasks. The ROC AUC score represents the probability that a randomly chosen positive edge $e^+$ is ranked higher than a randomly chosen negative edge $e^-$. A score of $0.5$ indicates random guessing, while $1.0$ indicates perfect separation.

Average precision (\textit{EPP}) summarizes the precision-recall curve into a single value. It is calculated as the weighted mean of precisions achieved at each threshold, where the weight is the increase in recall from the previous threshold. This metric is often preferred over AUC due to its robustness in imbalanced classes. Although in practice we used an equally sized positive and negative sample, the result is still meaningful.

\textit{EPRk} measures the proportion of positive edges recovered within the top $k\%$ of the ranked evaluation set $S$. Let $S_k$ denote the subset of edges in the top $k\%$ of the ranking. We report this metric for $k=5$ and $k=20$.

\begin{equation}
    {\rm EPRk} = \frac{|S_k \cap E^+|}{|S_k|}
    \label{eq:GRscoreDef3}
\end{equation}

\subsection*{Flywire Dataset}

Our study relied on the FlyWire FAFB data set \cite{dorkenwald2022flywire,dorkenwald2024neuronal}, a dense, whole-brain electron microscopy (EM) reconstruction of a female adult Drosophila melanogaster brain. The FlyWire project represents a large-scale, community-driven effort to segment, proofread, and curate the Female Adult Fly Brain (FAFB) EM volume, resulting in a high resolution, neuron-level connectome of the fly central nervous system. The dataset provides detailed morphological reconstructions of individual neurons and precise annotations of synaptic contacts, enabling the construction of a directed, weighted network that captures connectivity at single-synapse resolution.

For our experiments, we utilize the neuron cell-type annotations provided by the FlyWire community, which organize neurons into hierarchical classes based on their functional roles, anatomical location, and morphological characteristics (see Table~\ref{tab:flywire_stats} for more details of their frequency). This hierarchical labelling allows us to evaluate the embedding quality not only in terms of global structural reconstruction, but also in terms of the preservation of biologically meaningful organization across multiple levels of granularity.

The complete data set contains 134,181 neurons connected by 3,869,878 synapses. During preprocessing, we merge multi-edges corresponding to multiple synapses between the same ordered pair of neurons and remove neurons that are disconnected from the largest connected component. After this cleaning procedure, the resulting network consists of 132,483 nodes and 2,509,503 edges.
We provide a brief description for each super class:
\begin{itemize}
    \item \textbf{Central} "Local" interneurons that act as the brain's internal wiring. They are completely contained within the central brain complex and do not extend to the eyes, the nerve cord, or the rest of the body.
    \item \textbf{Ascending} Bottom-up inputs traveling from the ventral nerve cord (the insect equivalent of a spinal cord) up into the brain. These typically carry sensory feedback from the body or pre-processed signals from body segments.
    \item \textbf{Descending} Top-down outputs traveling from the brain down to the ventral nerve cord. These act as command signals, directing the body to move, walk, or fly.
    \item \textbf{Motor} Neurons that bypass the nerve cord and send axons directly from the brain to muscles in the periphery (usually controlling head or mouth movements).
    \item \textbf{Optic} Local processing neurons confined entirely within the visual centers (optic lobes or ocellar ganglion). They process visual data locally—much like a retina—before it is sent to the main brain.
    \item \textbf{Visual Projection} The output of the eyes; these neurons carry processed visual features from the optic lobes inward to the central brain for decision-making.
    \item \textbf{Visual Centrifugal} The feedback loop of the visual system; these neurons carry signals from the central brain back out to the optic lobes, to modulate attention or adjust visual sensitivity.
    \item \textbf{Sensory} Primary input neurons that bring raw data (such as smell, taste, or touch) directly from external sensors on the head into the brain for processing.
\end{itemize}

\begin{table}[h]
    \centering
    \begin{tabular}{lrr}
        \toprule
        \textbf{Super-Class} & \textbf{Count} & \textbf{Percentage (\%)} \\
        \midrule
        Optic Neurons        & 76,765         & 57.94 \\
        Central Neurons      & 32,298         & 24.38 \\
        Sensory Neurons      & 11,415         & 8.62 \\
        Visual Projection Neurons & 8,046     & 6.07 \\
        Ascending Neurons    & 1,989          & 1.50 \\
        Descending Neurons   & 1,276          & 0.96 \\
        Visual Centrifugal   & 519            & 0.39 \\
        Motor Neurons        & 106            & 0.08 \\
        Endocrine Neurons    & 69             & 0.05 \\
        \midrule
        \textbf{Total}       & \textbf{132,483} & \textbf{100.0} \\
    \end{tabular}
    \caption{\textbf{Neuron super-class sizes in the Flywire FAFB Dataset.} 
    The table shows the number and relative percentage of neurons belonging to each functional super-class in the Flywire FAFB Dataset\cite{FlyBrain_base_Nature}.}
    \label{tab:flywire_stats}
\end{table}

\subsection*{Embedding methods}

\subsubsection*{CLOVE}

CLOVE is an network embedding method that maps nodes of a given graph into the 2 dimensional hyperbolic space~\cite{CLOVE}. It works by constructing a multi-level angular arrangement of communities and sub-communities, refined recursively down to the level of individual nodes. Angular coordinates follow this iterative hierarchical modular structure, while radial coordinates are assigned using the Popularity-Similarity Optimization (PSO) model~\cite{PSO, our_d_dim_PSO, coalescentEmbedding, commSector_hypEmbBasedOnComms_2019}.

At the top level of the hierarchy ($l=0$), a non-overlapping community structure is extracted using a modularity-based detection method (CLOVE uses Leiden by default). The resulting communities $t_m^{(0)}$ form the super-nodes of a complete weighted \emph{proximity graph}. The weight between communities $i$ and $j$ is defined as
\begin{equation}
W_{ij} = \exp\left(\frac{2 E_l C_{ij}}{K_i K_j}\right) + 1,
\label{eq:clove_preweight}
\end{equation}
where $E_l = E_0 = E$ denotes the total number of edges in the network, $K_i$ and $K_j$ are the intra-community edge counts, and $C_{ij}$ is the number of edges connecting the two communities. Subsequently, a minimal-weight Hamiltonian cycle of the proximity graph, computed using a Christofides-based \textit{Travelling Salesman Problem} (TSP) solver with optional threshold-accepting refinement, determines the angular ordering of top-level communities. Each community $t_i^{(0)}$ is assigned a circular sector proportional to the number of nodes it encloses, i.e.,
\begin{equation}
\left[\Phi^{(0)}_{i,\mathrm{start}}, \Phi^{(0)}_{i,\mathrm{end}}\right)
=
\left[
\frac{2\pi}{N}\sum_{j=1}^{i-1} n_j^{(0)},
\;
\frac{2\pi}{N}\sum_{j=1}^{i} n_j^{(0)}
\right),    
\end{equation}
with $n_j^{(0)}$ denoting the number of nodes in community $t_j^{(0)}$.

At subsequent levels ($l+1$), each community $t_i^{(l)}$ is subdivided by running the same community detection method on its induced subgraph leading to sub-modules denoted by $t^{(l+1)}_{i1},\dots,t^{(l+1)}_{ik}$. For each parent community $t_i^{(l)}$, we construct a complete weighted graph using the same definition of $W_{ij}$ as in Eq.~(\ref{eq:clove_preweight}), but now also include its two neighbouring communities from the parental level $l$ as ''anchor'' nodes. Solving a specific local TSP on this augmented graph yields an ordering of the sub-modules that aligns with the parent’s angular position. Then, each sub-community $t_{ik}^{(l+1)}$ receives an angular segment inside its parent's angular interval as follows
\begin{equation}
\left[\Phi^{(l+1)}_{ik,\mathrm{start}}, \Phi^{(l+1)}_{ik,\mathrm{end}}\right)
=
\left[
\Phi^{(l)}_{i,\mathrm{start}}
+ \frac{2\pi}{N}\sum_{j=1}^{k-1} n^{(l+1)}_{ij},
\quad
\Phi^{(l)}_{i,\mathrm{start}}
+ \frac{2\pi}{N}\sum_{j=1}^{k} n^{(l+1)}_{ij}
\right),
\end{equation}
where $n^{(l+1)}_{ij}$ is the size of the corresponding sub-module. 

The previous recursive subdivision continues until no further meaningful partitioning is possible. At the lowest level, individual nodes are arranged equidistantly with an angular spacing of $2\pi/N$. By default, their ordering is determined using a degree-based heuristic: the highest-degree nodes within each lowest-level community are positioned at the center of the corresponding angular sector, while the remaining nodes are inserted iteratively and symmetrically, prioritizing connections to nodes that have already been placed. For further details of the CLOVE algorithm, see Ref.~\cite{CLOVE}.

Finally, the radial coordinates of the nodes are assigned independently of their angular positions, following the PSO model. More specifically, nodes are first sorted in decreasing order of their degree, and the $i$-th node is assigned a radial coordinate given by
\begin{equation}
r_i = \frac{2}{\zeta} \ln\left(\frac{N}{i}\right),
\end{equation}
with $\zeta = 1$ for the native disk representation. The popularity-fading parameter $\beta$ is determined from the degree distribution exponent $\gamma$ as
\begin{equation}
\beta = \frac{1}{\gamma - 1},
\end{equation}
where $\gamma$ is estimated empirically by fitting a power-law $p(k) \sim k^{-\gamma}$ to the observed degree distribution.

The exact time complexity of CLOVE is quite challenging to characterize, as it strongly depends on the structure of the multilevel dendrogram that is constructed inherently by the algorithm. In order to obtain an approximate bound, the community dendrogram can be idealized as a $b$-ary tree, where each level $l = 1, \ldots, \log_b(N) - 1$ consists of $b^l$ communities, each of size $N / b^l$. Assuming additionally that $b(N) \sim N^c$ for some $0 < c < 1$, the overall computational complexity is bounded above by
$\mathcal{O}(N^{2c+1})$ as shown in Ref.~\cite{CLOVE}. Indeed, numerical evaluations across a wide range of network types confirm this approximation, demonstrating that in practice the computational complexity lies between linear and quadratic scaling~\cite{CLOVE}.

\subsubsection*{Node2vec}

Node2vec is a network embedding method that maps nodes of a given graph into the $d$-dimensional Euclidean space while preserving structural properties of the network topology~\cite{node2vec}. The method is based on the assumption that the structural role and neighbourhood of a node can be effectively captured by exploring the network through random walks. To this end, Node2vec characterizes the local environment of each node by generating multiple truncated random walks that follow the network edges and produce ordered sequences of nodes with a fixed length~\cite{node2vec}.

These node sequences are treated analogously to sentences in natural language processing. More specifically, Node2vec adopts the same methodology as the \textit{Word2vec} algorithm~\cite{word2vec}, in which words that appear in similar contexts are embedded close to each other in the learned vector space. In Node2vec, sequences of nodes obtained from random walks replace textual sentences, and context windows of consecutive nodes replace word contexts. A neural network model is then trained to maximize the likelihood of observing neighbouring nodes within a given window, resulting in embeddings where nodes that frequently co-occur along random walks are placed at small Euclidean distances.

A key feature of Node2vec is the use of biased random walks, which allows for an interpolation between local and global exploration of the network~\cite{node2vec}. The transition probabilities of the random walkers are controlled by two parameters; first, the return parameter $p$ and, then a so-called in--out parameter $q$. Assuming a walk that has just moved from node $t$ to node $v$, the un-normalised transition probability to a neighbouring node $x$ is given by
\begin{equation}
\alpha_{pq}(t,x) =
\begin{cases}
\frac{1}{p}, & \text{if } x = t, \\
1, & \text{if } x \text{ is a common neighbour of } t \text{ and } v, \\
\frac{1}{q}, & \text{otherwise}.
\end{cases}    
\end{equation}
Low values of $p$ combined with high values of $q$ bias random walks toward the immediate neighbourhood of the starting node, thus favouring embeddings that mostly preserve the local connectivity structure. In contrast, high values of $p$ and low values of $q$ promotes the exploration of more distant nodes in the network, yielding embeddings that more strongly reflect the global topological organization. Consequently, tuning these parameters carefully enables Node2vec to control the trade-off between local and global structural information captured in the learned node representations. The computational complexity of Node2vec for embedding a graph with $N$ nodes and $E$ edges into a $d$-dimensional space is $
\mathcal{O}\left(E + N \cdot d \cdot \omega^2\right)$,
where $\omega$ denotes the context window size~\cite{node2vec}.

\section*{Data availability}
All data used in this work is publicly available from
\url{https://flywire.ai/}.

\section*{Code availability}
The Python implementation of CLOVE is publicly available at \url{http://github.com/samu32ELTE/hypCLOVE}, whereas Nod2vec is publicly available at \url{https://pypi.org/project/node2vec/}.



\section*{Acknowledgements}

This project has received funding from the European Union’s Horizon 2020 research and innovation programme under grant agreement no. 101021607
and was partially supported by the Data-Driven Health Division of National Laboratory for Health Security, Health Services Management Training Centre, Semmelweis University, Budapest, Hungary.

\section*{Author contributions}
G.P. developed the concept of the study; B.S. and S.G.B. implemented the methods; B.S. carried out the numerical simulations and performed the data analysis; B.S. prepared the figures; G.P., S.G.B. and B.S. participated in the writing of the manuscript and contributed to the interpretations of the results. All authors reviewed the manuscript.

\section*{Competing interests}
The authors declare no competing interests.

\clearpage

\begin{flushleft} 
{\LARGE \bfseries{SUPPLEMENTARY INFORMATION}}
\end{flushleft}

\setcounter{section}{0}
\setcounter{equation}{0}
\setcounter{figure}{0}
\setcounter{table}{0}
\renewcommand{\thesection}{S\arabic{section}}
\renewcommand{\thefigure}{S\arabic{figure}}
\renewcommand{\thetable}{S\arabic{table}}
\renewcommand{\theequation}{S\arabic{equation}}

\section{Supplementary Note 1}

\subsection{Additional superclass visualizations}

\begin{figure}[htbp]
    \centering

    \begin{subfigure}[t]{0.32\textwidth}
        \caption{\label{fig:subfigA2}}
        \includegraphics[width=\textwidth]{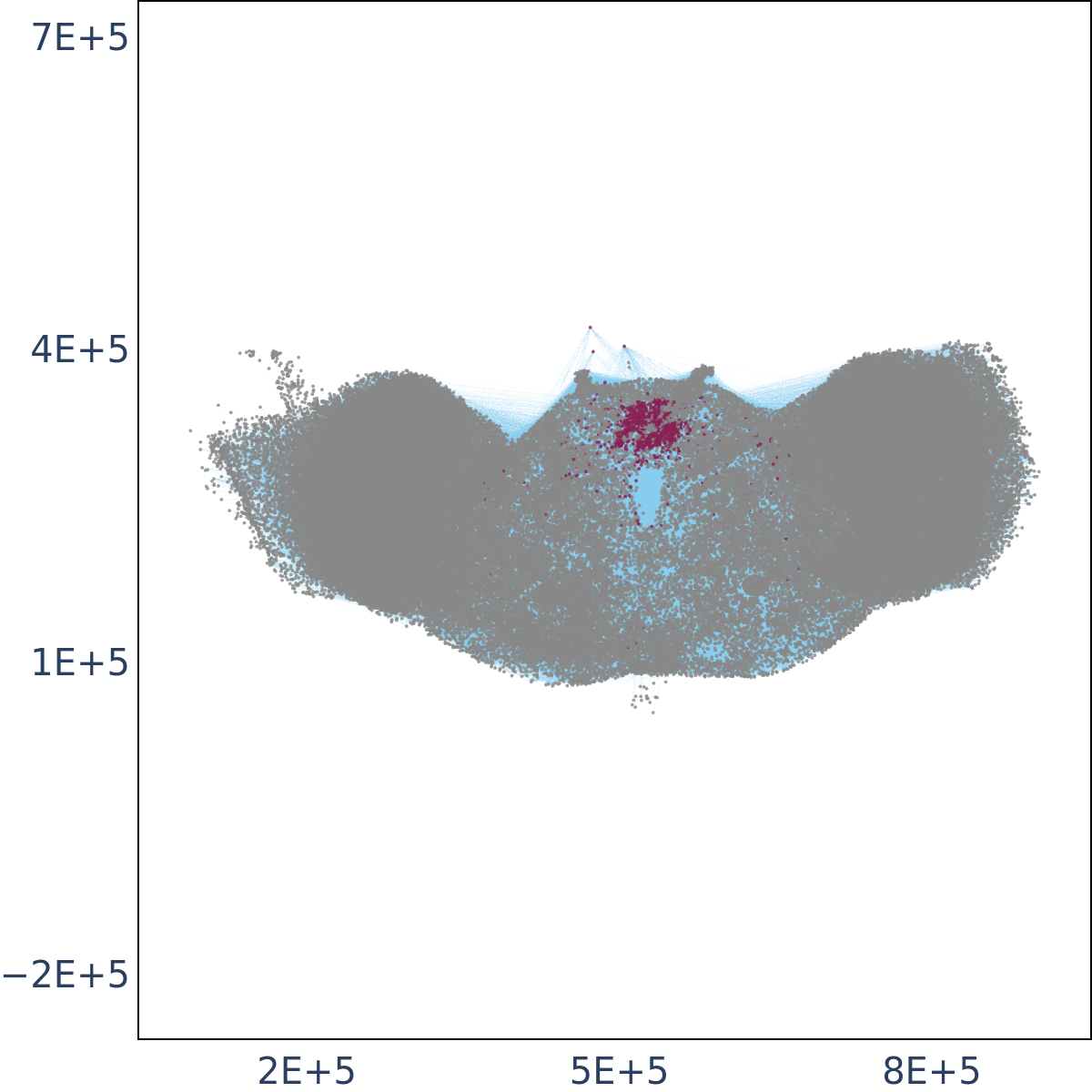}
    \end{subfigure}
    \begin{subfigure}[t]{0.32\textwidth}
        \caption{\label{fig:subfigB2}}
        \includegraphics[width=\textwidth]{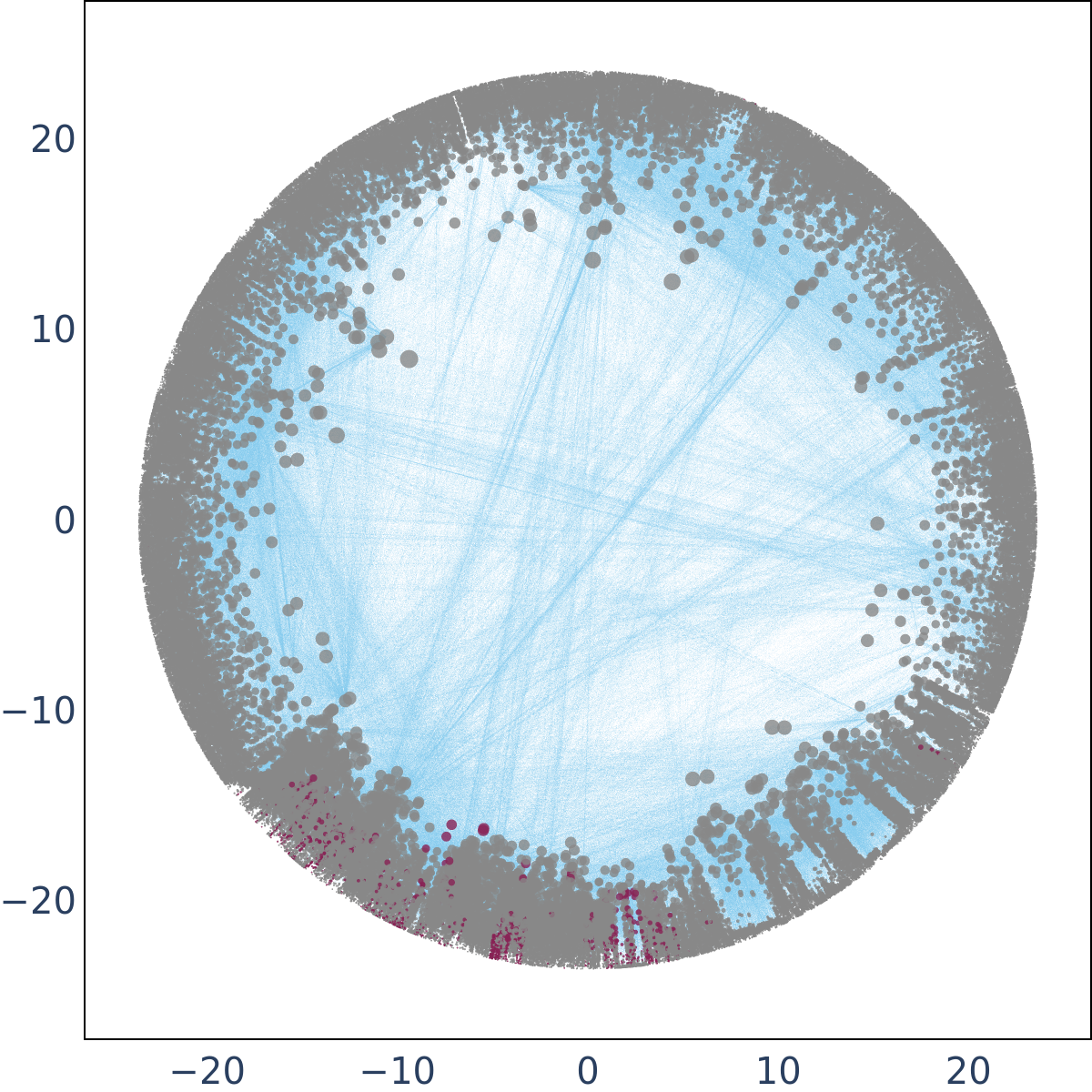}
    \end{subfigure}
    \begin{subfigure}[t]{0.32\textwidth}
        \caption{\label{fig:subfigC2}}
        \includegraphics[width=\textwidth]{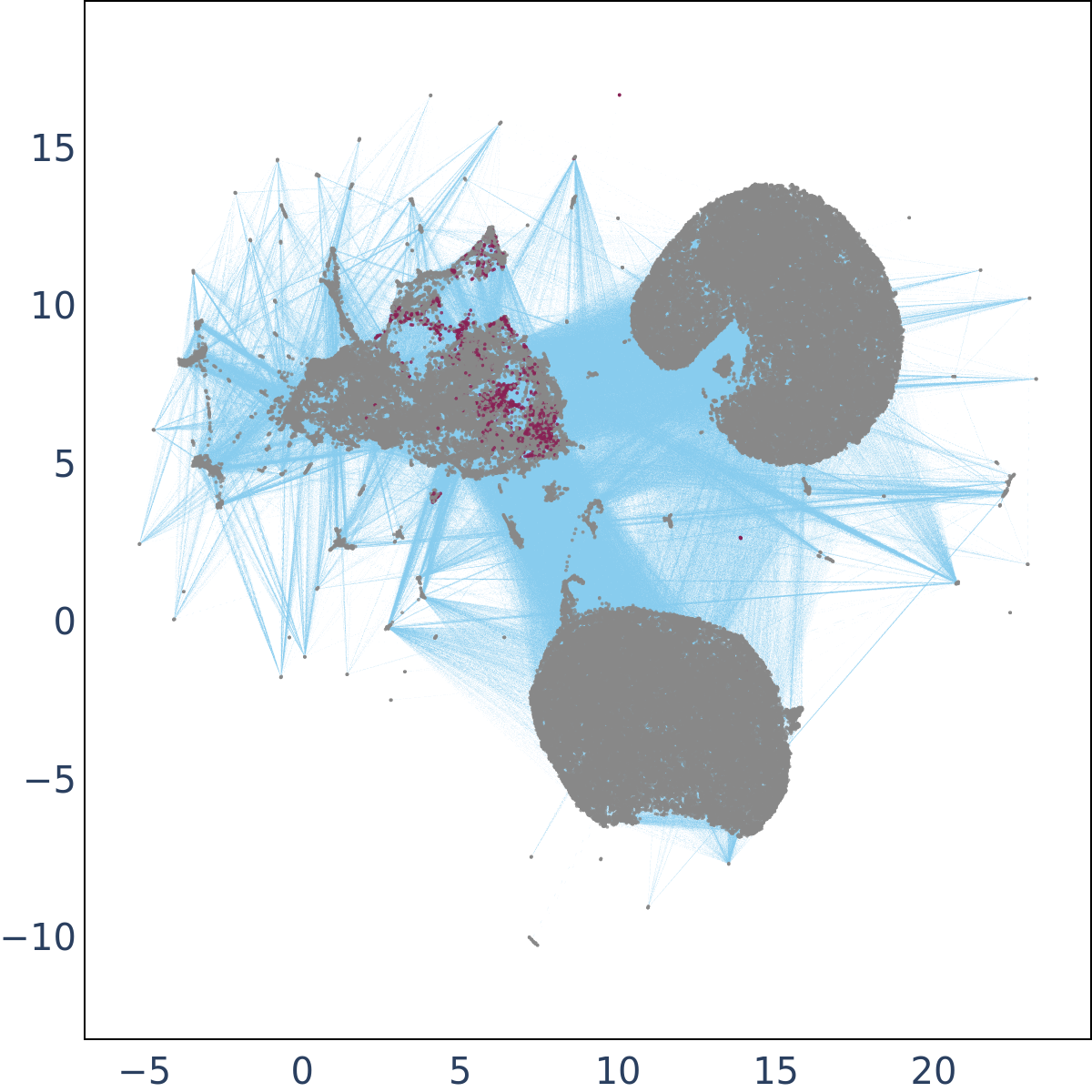}
    \end{subfigure}

    \caption{\textbf{Embeddings of the Drosophila brain network highlighting the "ascending" super class.}
    In panel a) we show a 2d Cartesian projection of the original neuron coordinates in the 3d Euclidean space.
    Panel b) displays the Hyperbolic embedding of the network in the 2d native disk representation according to CLOVE.
    In panel c) we display the 64d Euclidean embedding with node2vec, projected to 2d using UMAP.}
    \label{fig:embedd_illustr_ascending}
\end{figure}

\begin{figure}[htbp]
    \centering

    \begin{subfigure}[t]{0.32\textwidth}
        \caption{\label{fig:subfigA3}}
        \includegraphics[width=\textwidth]{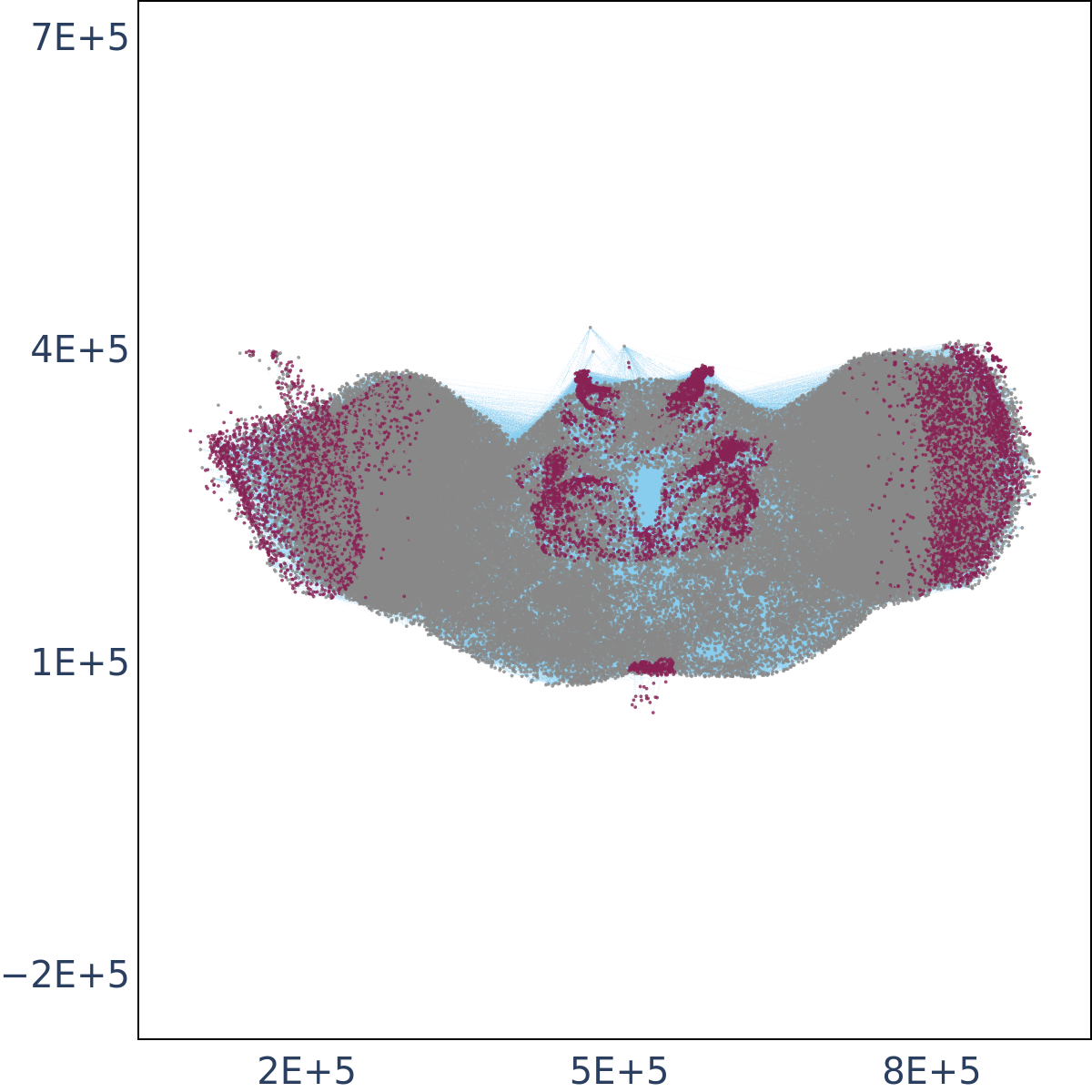}
    \end{subfigure}
    \begin{subfigure}[t]{0.32\textwidth}
        \caption{\label{fig:subfigB3}}
        \includegraphics[width=\textwidth]{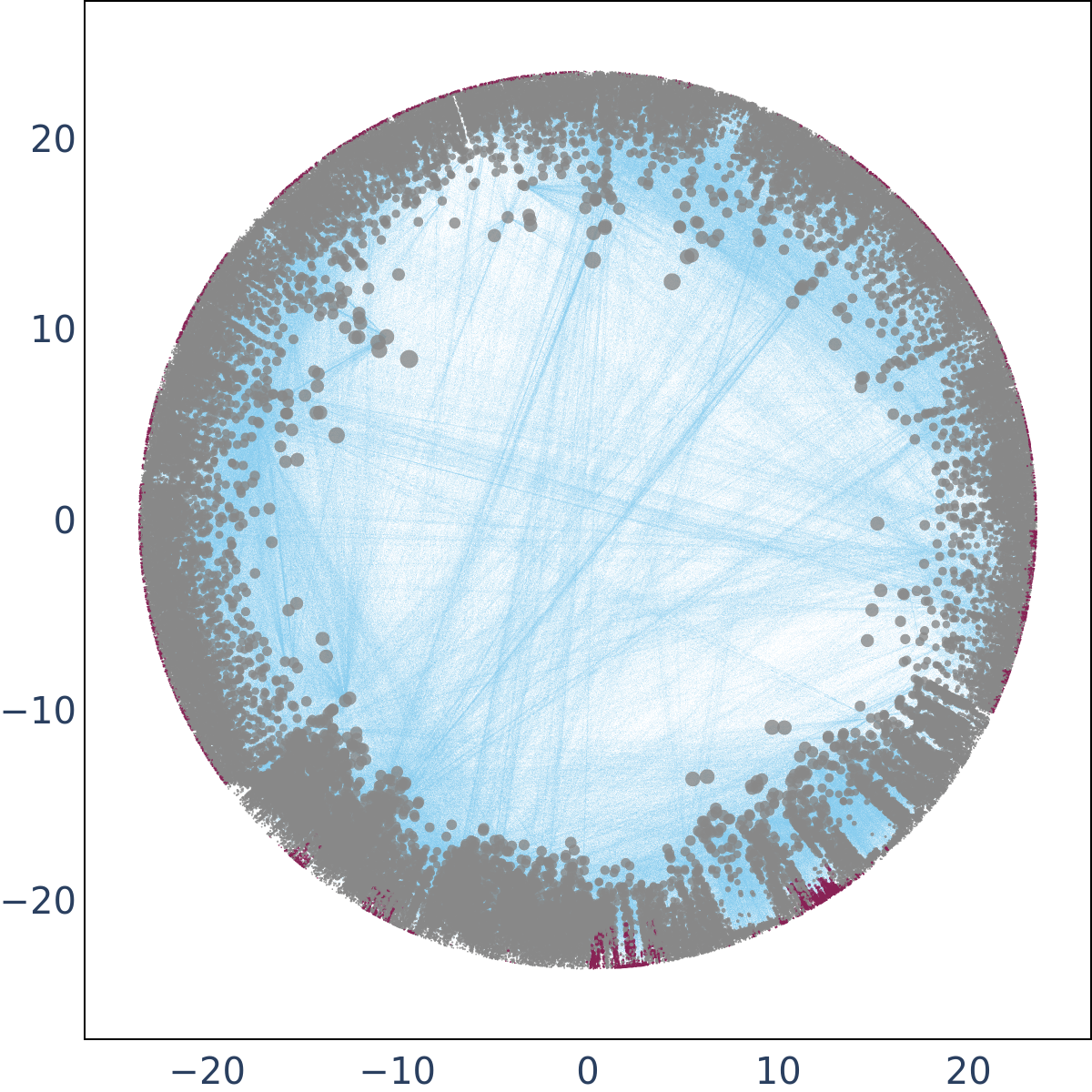}
    \end{subfigure}
    \begin{subfigure}[t]{0.32\textwidth}
        \caption{\label{fig:subfigC3}}
        \includegraphics[width=\textwidth]{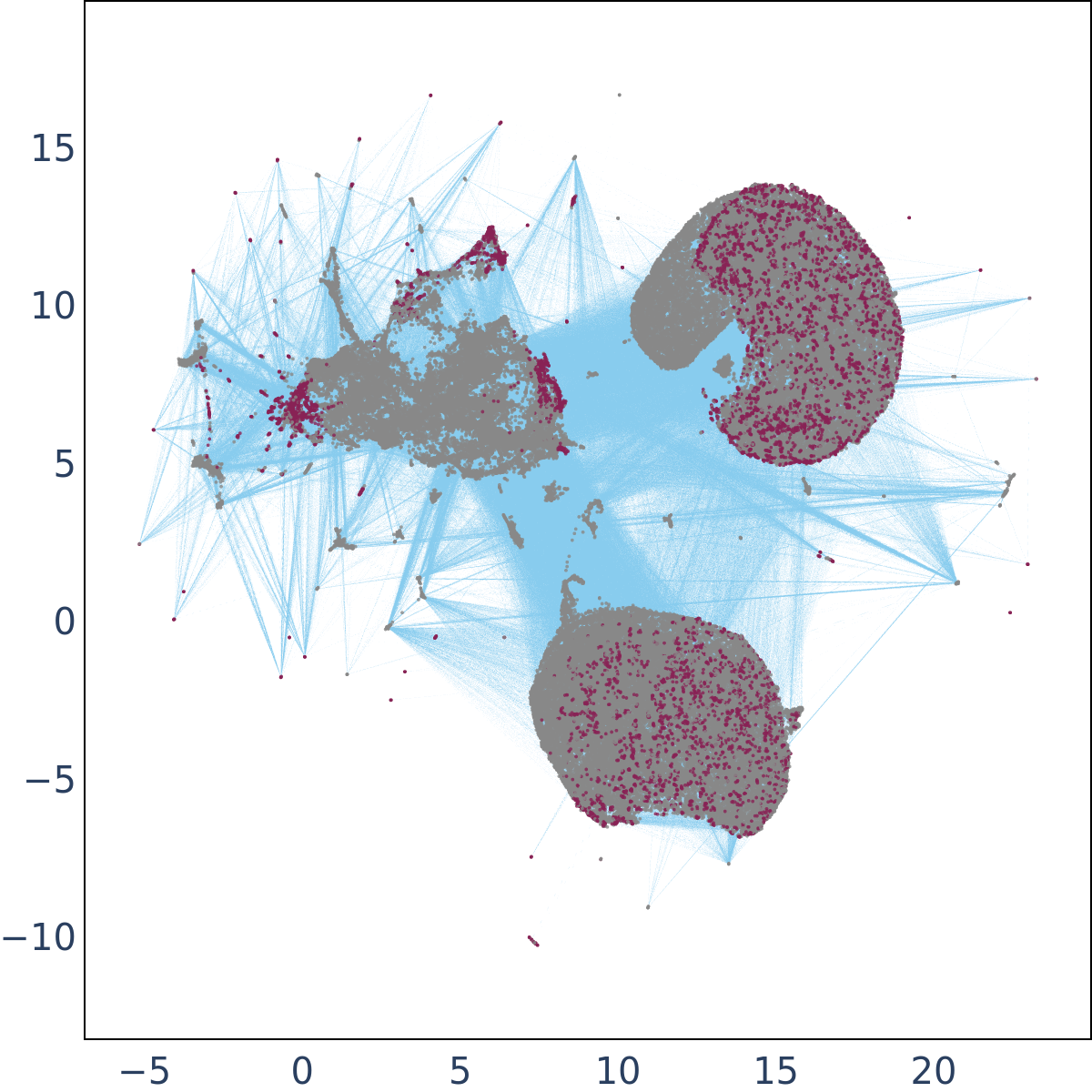}
    \end{subfigure}

    \caption{\textbf{Embeddings of the Drosophila brain network highlighting the "sensory" super class.}
    In panel a) we show a 2d Cartesian projection of the original neuron coordinates in the 3d Euclidean space.
    Panel b) displays the Hyperbolic embedding of the network in the 2d native disk representation according to CLOVE.
    In panel c) we display the 64d Euclidean embedding with node2vec, projected to 2d using UMAP.}
    \label{fig:embedd_illustr_sensory}
\end{figure}

\begin{figure}[htbp]
    \centering

    \begin{subfigure}[t]{0.32\textwidth}
        \caption{\label{fig:subfigA4}}
        \includegraphics[width=\textwidth]{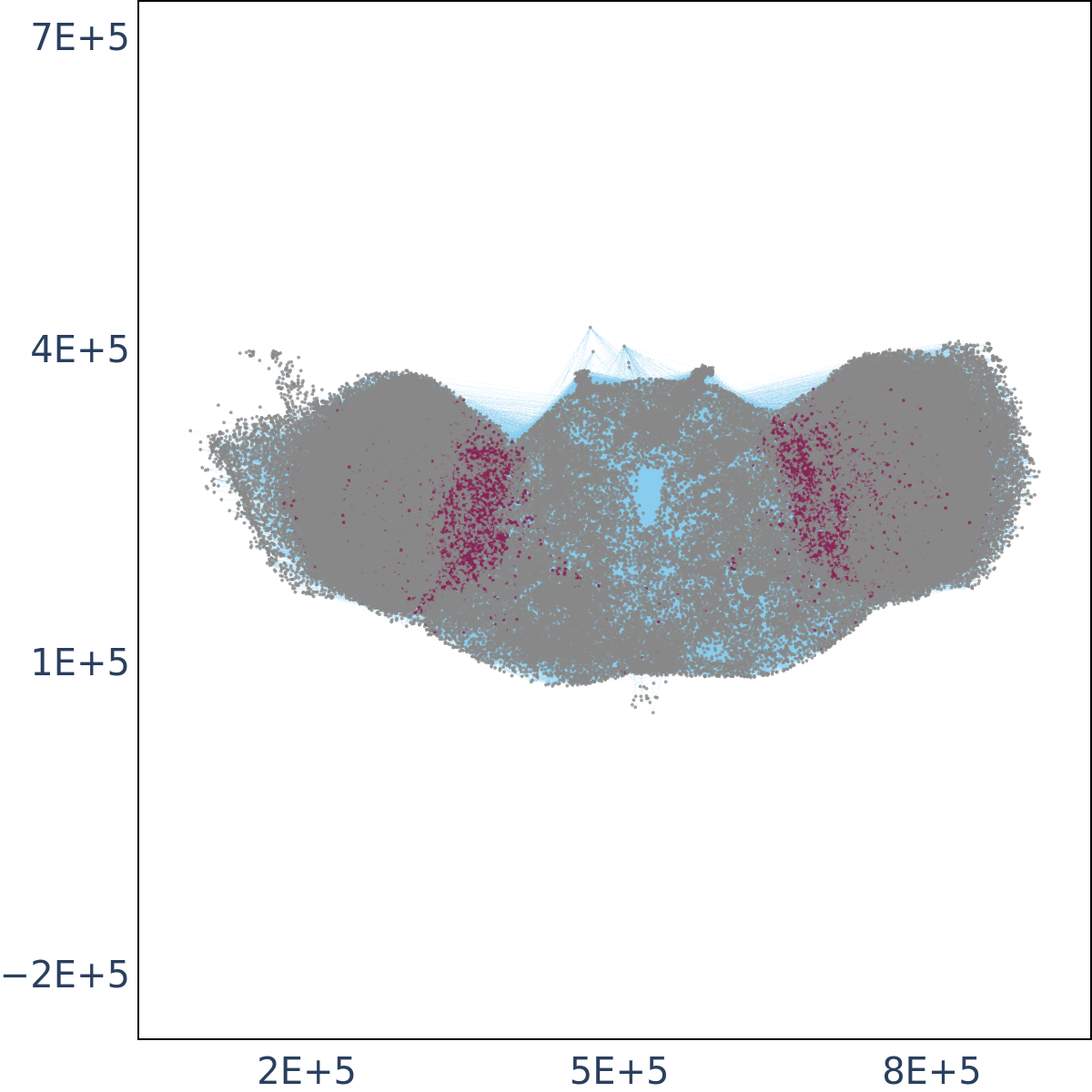}
    \end{subfigure}
    \begin{subfigure}[t]{0.32\textwidth}
        \caption{\label{fig:subfigB4}}
        \includegraphics[width=\textwidth]{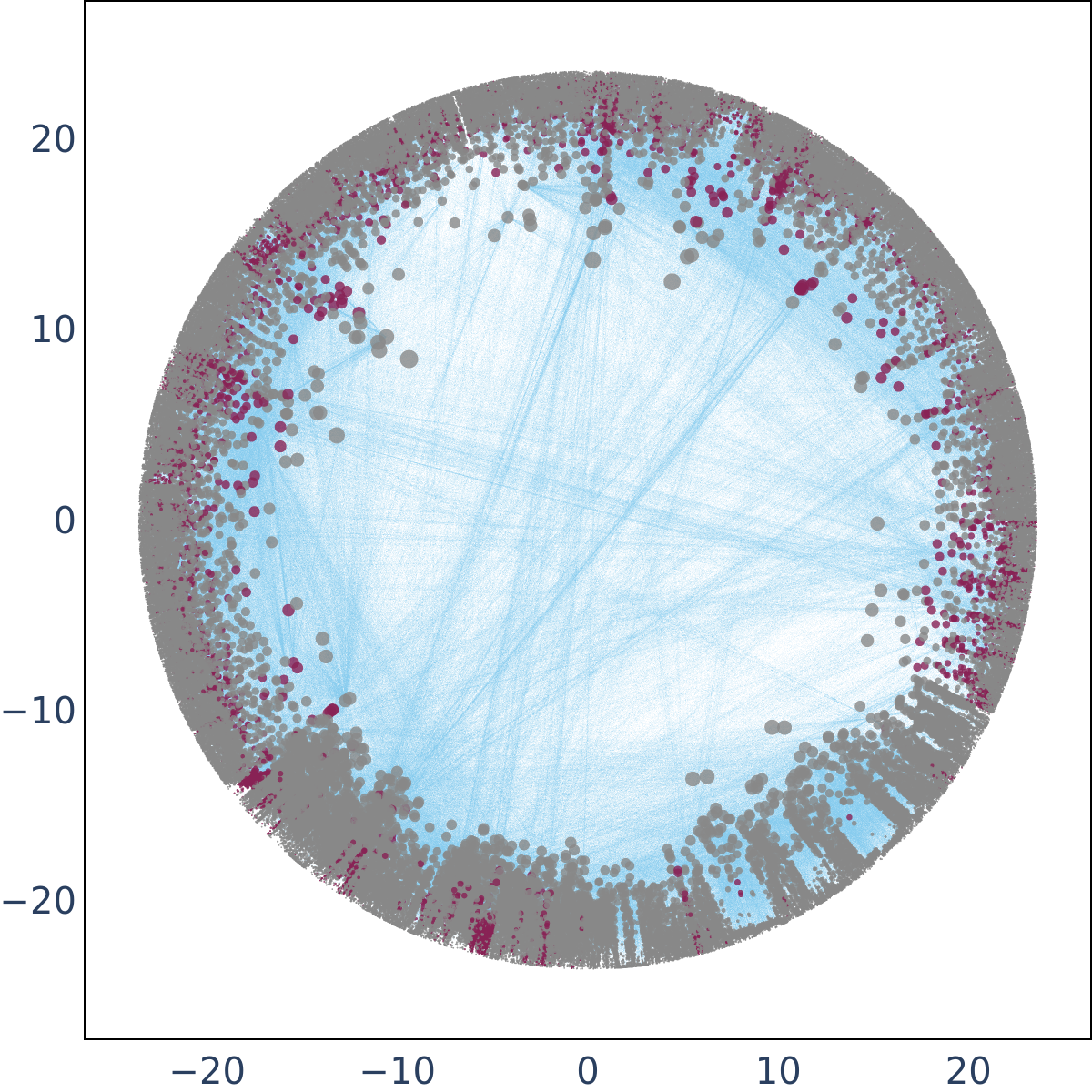}
    \end{subfigure}
    \begin{subfigure}[t]{0.32\textwidth}
        \caption{\label{fig:subfigC4}}
        \includegraphics[width=\textwidth]{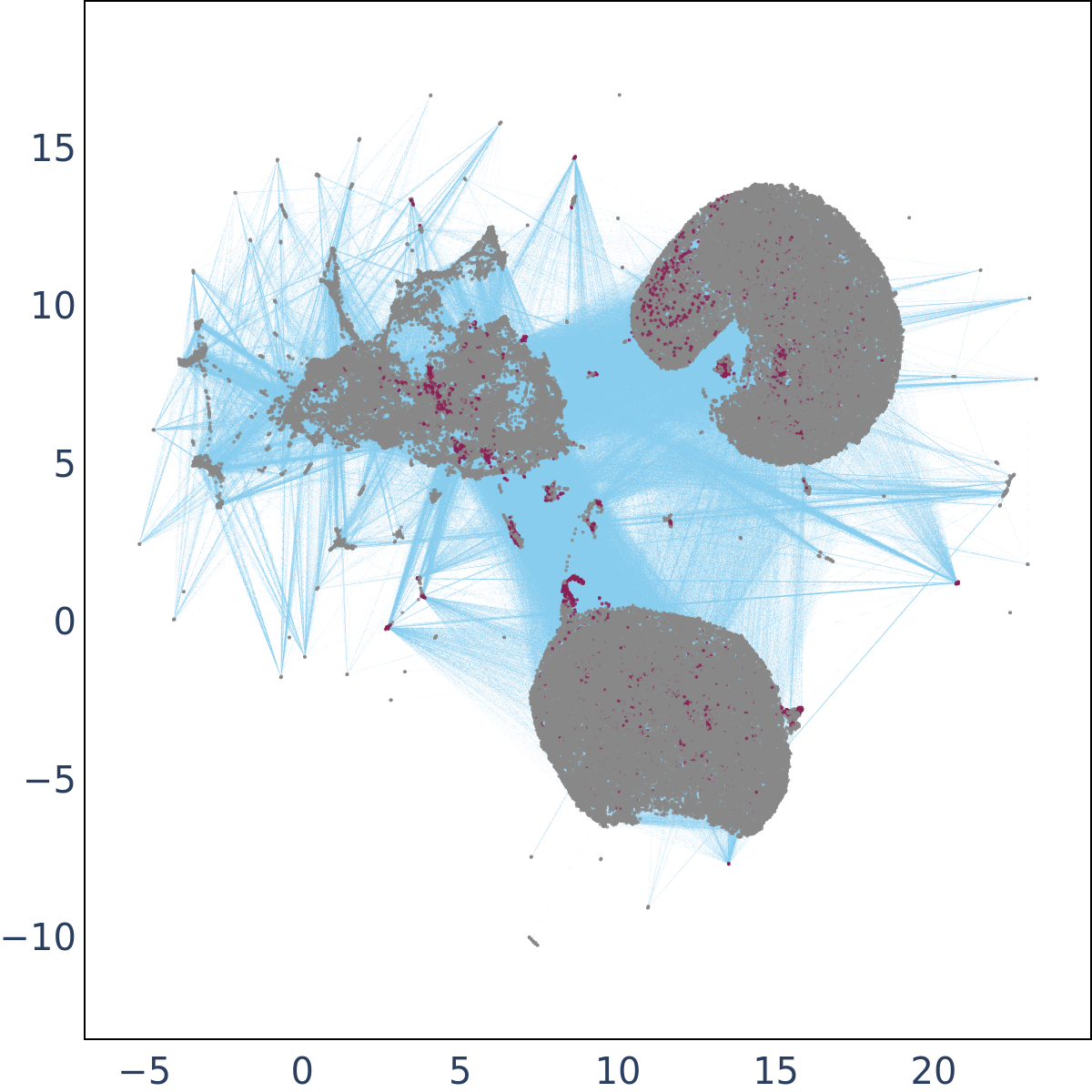}
    \end{subfigure}

    \caption{\textbf{Embeddings of the Drosophila brain network highlighting "visual projection" super class.}
    In panel a) we show a 2d Cartesian projection of the original neuron coordinates in the 3d Euclidean space.
    Panel b) displays the Hyperbolic embedding of the network in the 2d native disk representation according to CLOVE.
    In panel c) we display the 64d Euclidean embedding with node2vec, projected to 2d using UMAP.}
    \label{fig:embedd_illustr_visproj}
\end{figure}

\begin{figure}[htbp]
    \centering

    \begin{subfigure}[t]{0.32\textwidth}
        \caption{\label{fig:subfigA5}}
        \includegraphics[width=\textwidth]{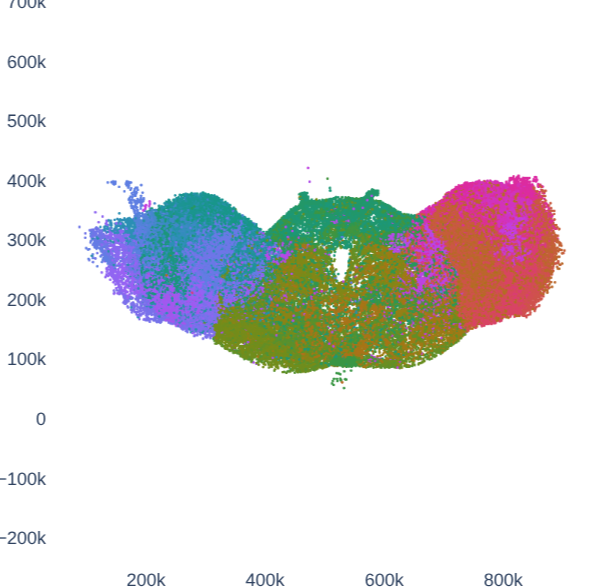}
    \end{subfigure}
    \begin{subfigure}[t]{0.32\textwidth}
        \caption{\label{fig:subfigB5}}
        \includegraphics[width=\textwidth]{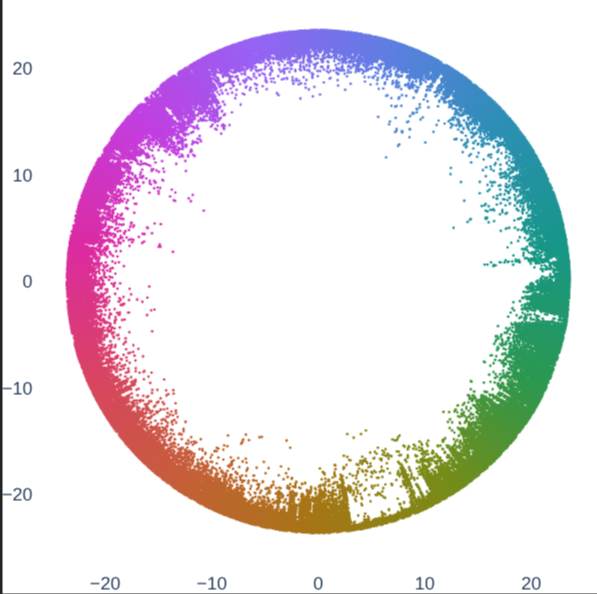}
    \end{subfigure}
    \begin{subfigure}[t]{0.32\textwidth}
        \caption{\label{fig:subfigC5}}
        \includegraphics[width=\textwidth]{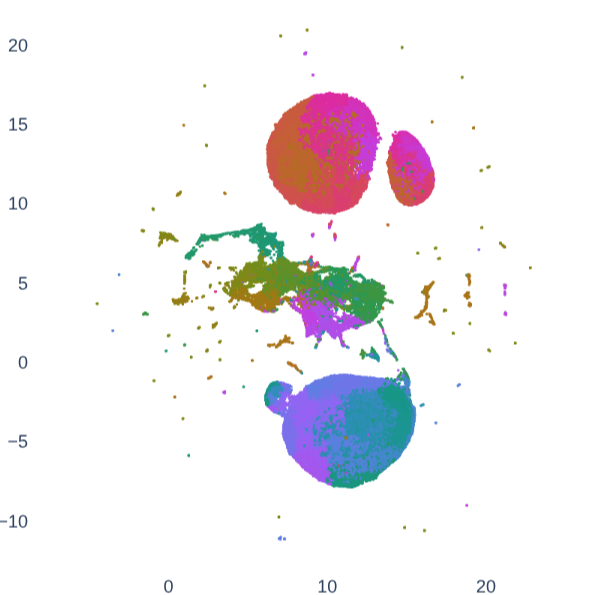}
    \end{subfigure}

    \caption{\textbf{Embeddings of the Drosophila brain network highlighting the hyperbolic angular arrangement.}
    In panel a) we show a 2d Cartesian projection of the original neuron coordinates in the 3d Euclidean space.
    Panel b) displays the Hyperbolic embedding of the network in the 2d native disk representation according to CLOVE.
    In panel c) we display the 64d Euclidean embedding with node2vec, projected to 2d using UMAP. The node colours are defined based on the angular coordinate in panel b).}
    \label{fig:embedd_illustr_rainbow}
\end{figure}

We show additional layouts of the the Drosophila brain network in Figs.~\ref{fig:embedd_illustr_ascending}-\ref{fig:embedd_illustr_rainbow}. In each Figure, a different node colouring is applied, while the arrangement of the nodes is identical with that in Fig.~1 in the main paper. In Fig.~\ref{fig:embedd_illustr_ascending}. we highlighted the "ascending" super class of the neurons with red node colour, while the rest of the nodes remained gray. Similarly, in Fig.~\ref{fig:embedd_illustr_sensory}. we highlighted the "sensory" super class in the same fashion, whereas in Fig.~\ref{fig:embedd_illustr_visproj}. it is the "visual projection" super class that received the red colour. According to Table 1. in the main paper, the remaining further four superclasses contain a very small amount of neurons compared to the entire system, therefore, these are not visualised here.

In Fig.~\ref{fig:embedd_illustr_rainbow}. the nodes are coloured according to the angular coordinate in the hyperbolic disk. According to the figure, the quasi-1D manifold that represents the angular arrangement in hyperbolic space corresponds to a predominantly continuous manifold in the original 3D Euclidean space. Stated differently, the angular arrangement resulting from the hyperbolic projection of the original spatial layout is far from random. Rather, it resembles following a complex but roughly continuous spatial path through the neurons of the fly's brain.

\captionsetup[subfigure]{justification=raggedright,singlelinecheck=off}

\captionsetup[subfigure]{justification=raggedright,singlelinecheck=off}

\captionsetup[subfigure]{justification=raggedright,singlelinecheck=off}

\captionsetup[subfigure]{justification=raggedright,singlelinecheck=off}

\bibliographystyle{sn-nature}
\bibliography{fly_geom} 

\end{document}